\documentclass[aps,prd,a4paper,onecolumn,amsmath,preprintnumbers,notitlepage]{revtex4-1}

\usepackage{verbatim}
\usepackage[T1]{fontenc}
\usepackage[utf8]{inputenc}
\usepackage[american]{babel}
\usepackage{epsfig}
\usepackage{graphicx}
\usepackage{booktabs}
\usepackage{multirow}
\usepackage{dcolumn}
\usepackage{amsmath}
\usepackage{mathtools}
\usepackage{amsfonts}
\usepackage{amssymb}
\usepackage{epstopdf}
\usepackage{bm}
\usepackage{siunitx}
\usepackage{braket}
\usepackage{enumitem}
\usepackage{soul}
\usepackage[table]{xcolor}
\usepackage{color}
\usepackage{transparent}
\usepackage{pifont}

\definecolor{navyblue}{rgb}{0.0, 0.0, 0.5}
\definecolor{royalblue}{rgb}{0.25, 0.41, 0.88}
\definecolor{cadmiumgreen}{rgb}{0.0, 0.42, 0.24}
\definecolor{blue-violet}{rgb}{0.54, 0.17, 0.89}
\definecolor{darkviolet}{rgb}{0.58, 0.0, 0.83}
\definecolor{orange(colorwheel)}{rgb}{1.0, 0.5, 0.0}

\usepackage{hyperref}
\hypersetup{
    colorlinks=true, 
    linkcolor=royalblue, 
    citecolor=magenta}

\newcommand\ee{\end{equation}}
\newcommand\be{\begin{equation}}
\newcommand\eea{\end{eqnarray}}
\newcommand\bea{\begin{eqnarray}}

\newcommand{\om}{\Omega_m}
\newcommand{\omk}{\Omega_k}
\newcommand{\omrad}{\Omega_r}
\newcommand{\sige}{\sigma_8} 
\newcommand{\fsig}{f\sigma_8}

\newcommand{\lcdm}{$\Lambda$CDM}

\usepackage{booktabs}
\usepackage{multirow}
\usepackage{dcolumn}
\usepackage{colortbl}

\definecolor{magenta(process)}{rgb}{1.0, 0.0, 0.56}

\definecolor{darkspringgreen}{rgb}{0.09, 0.45, 0.27}

\definecolor{royalblue(web)}{rgb}{0.25, 0.41, 0.88}

\begin{document}

\title{$H_0$ Ex Machina: Vacuum Metamorphosis and Beyond $H_0$} 

\author{Eleonora Di Valentino}
\email{eleonora.divalentino@manchester.ac.uk}
\affiliation{Jodrell Bank Center for Astrophysics, School of Physics and Astronomy, University of Manchester, Oxford Road, Manchester, M13 9PL, UK}
\author{Eric V.\ Linder}
\email{evlinder@lbl.gov}
\affiliation{Berkeley Center for Cosmological Physics \& Berkeley Lab, University of California, Berkeley, CA 94720, USA}
\affiliation{Energetic Cosmos Laboratory, Nazarbayev University, Astana, Kazakhstan 010000} 
\author{Alessandro Melchiorri}
\email{alessandro.melchiorri@roma1.infn.it}
\affiliation{Physics Department and INFN, Universit\`a di Roma ``La Sapienza'', Ple Aldo Moro 2, 00185, Rome, Italy} 

\date{\today}

\preprint{}
\begin{abstract}
We do not solve tensions with concordance cosmology; we do obtain 
$H_0\approx 74\,$km/s/Mpc from CMB+BAO+SN data in our model, but that is not the point.  Discrepancies in Hubble constant values obtained by various astrophysical 
probes should not be viewed in isolation. While one can resolve at least 
some of the differences through either an early time transition or late 
time transition in the expansion rate, these introduce other changes. We 
advocate a holistic approach, using a wide variety of cosmic data, rather 
than focusing on one number, $H_0$. Vacuum metamorphosis, a late time 
transition physically motivated by quantum gravitational effects and with 
the same number of parameters as \lcdm, can successfully give a high 
$H_0$ value from cosmic microwave background data but fails when combined 
with multiple distance probes. We also explore the influence of spatial 
curvature, and of a conjoined analysis of cosmic expansion and growth. 
\end{abstract}

\maketitle

\section{Introduction} \label{sec:intro} 

Considerable literature has been devoted to differences in values for the 
present cosmic expansion rate, $H_0$, obtained by local distance ladder 
measurements using Cepheid star calibrators \cite{Riess:2019cxk} or tip of the 
red giant branch stars \cite{Freedman:2020dne}, by the cosmic microwave background 
(CMB) data \cite{Aghanim:2018eyx,Bianchini:2019vxp}, by baryon acoustic oscillation data in 
conjunction with primordial nucleosynthesis constraints but independent 
of CMB \cite{Addison:2017fdm, Macaulay:2018fxi, Cuceu:2019for}, and by strong gravitational 
lensing time delays \cite{Wong:2019kwg}. We do not opine here on the possibility 
of inaccurate measurements or techniques (systematics) but instead pursue 
the avenue of actual new cosmic physics: the fault is not in our stars, but in ourselves (i.e.\ cosmological model). 

Two modifications of cosmological history have long been known as capable 
of raising the value determined by probes for the present expansion rate: 
an early time transition in the expansion rate (e.g.\ through adding extra 
energy density), thus decreasing the sound horizon 
\cite{Efstathiou:1998xx,Eisenstein:2004an,Doran:2006xp,Linder:2008nq,Hojjati:2013oya}, and a late time 
transition in the expansion rate (through either adding extra energy  
density or changing the Friedmann equation relating expansion rate to 
energy density), thus directly raising $H_0$ \cite{DiValentino:2017rcr,Li:2019yem,Pan:2019hac,Li:2020ybr,Khosravi:2017hfi,DiValentino:2020naf}. 

Each of these, however, have other cosmological effects besides changing 
$H_0$, e.g.\ on distances, the value of the matter density $\om$, CMB 
anisotropies, the amplitude or rate of growth of structure 
($\sige,\fsig,S_8$). For example, early time transitions must quickly shed 
their extra energy density so as not to disrupt the fit to CMB anisotropies 
or cosmic structure formation. 
Indeed, \cite{Hojjati:2013oya} actually detected such a transition up and down 
in the Planck 2013 and WMAP9 CMB data and showed the shift in $H_0$. 
Recently this idea has been revisited by 
\cite{Poulin:2018cxd,Smith:2019ihp,Agrawal:2019lmo,Lin:2019qug,Niedermann:2019olb}. 
Late time transitions that raise $H_0$ generally lower $\om$ (e.g.\ if 
keeping the well measured value of $\om h^2$), changing distances to 
sources, the growth of structure, and generally also the sound horizon 
and CMB anisotropies. Some of these can be compensated for with other 
changes but it is difficult to match all the data. 
Recent treatments of early time transitions \cite{Hill:2020osr} and late time transitions \cite{Benevento:2020fev} highlight some of the issues, while the overall situation is summarized in \cite{Knox:2019rjx}. There are a huge number of 
papers discussing specific aspects and models: 
we refer the reader to references in \cite{Knox:2019rjx} as well as more recent ones that evaluate viability with respect to multiple, diverse data sets \cite{DiValentino:2019ffd,Arendse:2019hev,Garcia-Quintero:2019cgt,Hart:2019dxi,DiValentino:2019jae,Liu:2019awo,Ivanov:2019hqk,Alcaniz:2019kah,Frusciante:2019puu,Yang:2020zuk,Jedamzik:2020krr,Pan:2020zza,Wu:2020nxz,Ye:2020btb,Braglia:2020iik,Blinov:2020hmc,Wang:2020zfv,Chudaykin:2020acu,Alestas:2020mvb,Clark:2020miy,Ballardini:2020iws,Keeley:2020rmo,Niedermann:2020dwg,Archidiacono:2020yey}. 

Here we address a late time transition in fundamental physics, arising 
from the well motivated quantum gravity effect of Parker's vacuum 
metamorphosis \cite{Parker:2000pr,Parker:2003as,Caldwell:2005xb}. This follows on 
the early attempt to use vacuum metamorphosis to attain higher $H_0$ in 
\cite{DiValentino:2017rcr}, which was successful for the probes considered. We 
emphasize that this is a first principles theory, not a phenomenological 
parametrization. A theory with a similar transition but different origin 
is {\"u}bergravity \cite{Khosravi:2017hfi}. We will go beyond these works by 
exploring the role of spatial curvature and adding further observational 
probes. 

Section~\ref{sec:vm} describes the vacuum metamorphosis theory in both  
its original and VEV forms, with the same and one more number of parameters 
as \lcdm, respectively. In Sec.~\ref{sec:data} we present the data sets we will use in different combinations. 
Section~\ref{sec:results} discusses the 
cosmological constraints including on $H_0$ and the 
spatial curvature, paying close attention to goodness of fit. We highlight in 
Sec.~\ref{sec:conjoin} the 
importance of using a wide range of different robust probes, and effects 
on cosmic structure growth, including a conjoined analysis. We discuss general lessons about late time 
transitions and conclude in  Sec.~\ref{sec:concl}.

\section{Vacuum Metamorphosis} \label{sec:vm} 

Vacuum metamorphosis arises from a nonperturbative summation of quantum 
gravity loop corrections due to a massive scalar field. The first order 
loop correction is familiar as Starobinsky $R^2$ gravity \cite{Starobinsky:1980te}, 
where $R$ is the Ricci scalar. When the Ricci scalar evolves during cosmic 
history to reach the scalar field mass squared, then a phase transition 
occurs and $R$ freezes with $R=m^2$. This changes the expansion rate from 
the earlier time, pure matter evolution. That is, at times before the phase 
transition the action is purely of the Einstein-Hilbert form without any 
dark energy, while  at later times the cosmic expansion has a different 
evolution. Other theories giving a phase transition in $R$ include Sakharov's 
induced gravity \cite{Sakharov:1967pk} and the sum over states approach of {\"u}bergravity 
\cite{Khosravi:2017hfi}. 

Vacuum metamorphosis is a highly predictive theory, as it has the same number 
of parameters as \lcdm. In the original form there is a relation between $m^2$ 
and the present matter density $\om$ (which also determines the transition 
redshift). Another possible form is where the 
massive scalar field has a vacuum expectation value (VEV) that manifests as 
a  cosmological constant at higher redshift -- thus the cosmology at earlier 
times than the transition is purely \lcdm. Here the VEV, or magnitude of 
the high redshift cosmological constant, is another free parameter. We will 
explore the constraints of the data on both the original and VEV forms of 
vacuum metamorphosis. 

We now review the key equations of vacuum metamorphosis (VM), extending them to  
include spatial curvature, and discussing our implementation of them. 
While we regard 
the original VM model without cosmological constant as the most 
elegant and theoretically compelling, we 
provide the general equations for the VEV form, with the original no high 
redshift cosmological 
constant form as a special case. 

The phase transition criticality condition is 
\be 
R=6(\dot H+2H^2+ka^{-2})=m^2 \ , 
\ee 
and, defining $M=m^2/(12H_0^2)$, the expansion behavior above and below 
the phase transition is 
\bea 
H^2/H_0^2&=&\Omega_m (1+z)^3+\Omega_r (1+z)^4 +\omk (1+z)^2 
+M\left\{1-\left[3\left(\frac{4}{3\Omega_m}\right)^4 M(1-M-\omk-\omrad)^3\right]^{-1}
\right\}, \ z>z_t \label{eq:habove}\\ 
H^2/H_0^2&=&(1-M-\omk)(1+z)^4+\omk(1+z)^2+M\, ,\quad z\le z_t \label{eq:hbelow} 
\eea 
where $\omk=-k/H_0^2$ is the spatial curvature effective energy density and $\omrad$ is 
the radiation energy density. 
The phase transition occurs at 
\be 
z_t=-1+\frac{3\Omega_m}{4(1-M-\omk-\omrad)} \ . \label{eq:zt}
\ee 

We see that above the phase transition, the universe behaves as one with 
matter (plus radiation plus spatial curvature) plus a cosmological constant, and after the phase transition it 
effectively has a radiation component (the matter and usual radiation is hidden within 
this expression) that rapidly redshifts away leaving a 
de Sitter phase. The original model did not include a VEV; we see that this 
lack of an explicit high redshift cosmological constant implies that 
\be 
\Omega_m=\frac{4}{3}\left[3M(1-M-\omk-\omrad)^3\right]^{1/4} \ . 
\qquad({\rm no\ VEV\ case}) \label{eq:vmom}
\ee 
So there is only one free parameter in the original model, either $M$ or $\Omega_m$, the same number as in \lcdm. 
For example, $\Omega_m=0.3$ implies $M=0.9017$, and $z_t=1.29$. We emphasize that the de Sitter 
behavior at late times is not a result of a cosmological constant, but 
rather the intrinsic physics of the model. 

The effective dark energy equation of state (i.e.\ of the effective 
component once the matter and normal radiation contributions have been accounted for) is 
\be 
w(z)=-1-\frac{1}{3}\frac{3\Omega_m (1+z)^{3}-4(1-M-\omk-\omrad)(1+z)^{4}}{M+(1-M-\omk-\omrad)(1+z)^{4}-\Omega_m (1+z)^{3}} \ , 
\ee 
below the phase transition, and simply $w(z>z_t)=-1$ above the phase 
transition. In the case without a cosmological constant (no VEV), there is no 
dark energy above the transition. 

The equation of state behavior is phantom, and more deeply phantom as 
the VEV diminishes. 
Note that for $M>0.9017$ (in the $\om=0.3$ case), the VEV can go negative, and 
this leads initially to a highly positive equation of state just after 
the transition. This is not an observationally viable region. As $M$ 
falls below the 
critical value, the VEV smooths out the rapid time 
variation, leading to a nearly constant $w(a)$. If $M$ falls too low, 
then the transition occurs in the future (see Eq.~\ref{eq:zt}), and we 
have simply the \lcdm\ model for the entire history to the present. 
Moreover, $M$ then becomes no longer a free parameter but is given in 
terms of $\om$ by the requirement that $H(z=0)/H_0=1$. Thus, when 
considering the VM VEV model one would need to place lower and upper limits on the prior of the extra free parameter, corresponding to 
$z_t\ge0$ and $\Omega_{\rm de}(z>z_t)\ge0$ respectively. 

The lower bound on $M$ from Eq.~(\ref{eq:zt}) is 
\be 
M_{\rm lower}=1-\frac{3\om}{4}-\omk-\omrad \ . 
\ee 
Determining the upper bound on $M$, i.e.\ the nonnegativity of the curly 
brackets in Eq.~(\ref{eq:habove}), requires solving a quartic equation. 
Therefore we instead choose $M$ as our free parameter and place the 
prior bounds on $\om$, for which there is an explicit analytic 
solution. In this case the bounds on $\om$ become 
\be 
\frac{4}{3}(1-M-\omk-\omrad) \le\om\le \frac{4}{3}\left[3M(1-M-\omk-\omrad)^3\right]^{1/4}\,, \label{eq:ombounds} 
\ee 
where the lower bound corresponds to the condition $z_t\ge0$ and the upper bound to 
$\Omega_{\rm de}(z>z_t)\ge0$.

\section{Data} \label{sec:data} 

In order to constrain the VM model parameters, we utilize various combinations of some of the most recent cosmological measurements available:

\begin{itemize}

\item {\bf CMB}: Temperature and polarization CMB angular power spectra of the Planck legacy release 2018 {\it plikTTTEEE+lowl+lowE}~\cite{Aghanim:2018eyx,Aghanim:2019ame}. This serves as our 
baseline data set and is 
included in all data combinations. 

\item {\bf CMB lensing}: CMB lensing reconstruction power spectrum data 2018, obtained with a CMB trispectrum analysis in~\cite{Aghanim:2018oex}.  
This data is only included in cases where it is specifically listed.  

\item {\bf BAO}: Baryon Acoustic Oscillation measurements 6dFGS~\cite{Beutler:2011hx}, SDSS MGS~\cite{Ross:2014qpa}, and BOSS DR12~\cite{Alam:2016hwk}, in the same combination used by the Planck collaboration in~\cite{Aghanim:2018eyx}.

\item {\bf SN}: Luminosity distance data of $1048$ Type Ia Supernovae from the Pantheon catalog~\cite{Scolnic:2017caz}.

\item {\bf R19}: Gaussian prior $H_0=74.03\pm1.42$ km/s/Mpc at 68\% CL on the Hubble constant as measured by the SH0ES collaboration in~\cite{Riess:2019cxk}.

\end{itemize}

We assume initially a 6-dimensional parameter space, varying at the same time the baryon energy density $\Omega_bh^2$, the ratio of the sound horizon at decoupling to the angular diameter distance to last scattering $\theta_{MC}$, the optical depth to reionization $\tau$, the amplitude and the spectral index of the primordial scalar perturbations $A_s$ and $n_s$, and the 
vacuum metamorphosis parameter $M$ defined in Sec.~\ref{sec:vm} and related to the matter density $\Omega_m$ through the Eq.~(\ref{eq:vmom}). 
A second set of analyses includes the curvature density $\Omega_k$, i.e.\ a spatially nonflat universe, as a seventh parameter. 
Each of these sets is then also analyzed  for the 
VM VEV model where one more degree of freedom is present, i.e.\  relaxing the condition of Eq.~(\ref{eq:vmom}), and also allowing the cold dark matter density $\Omega_ch^2$, equivalent to $\om$ independent of 
$M$, to vary. We use flat uniform priors on these parameters, as reported in Table~\ref{tab:priors}.

In order to study the data and evaluate the constraints on the cosmological parameters, we use our modified version of the publicly available Monte-Carlo Markov Chain package \texttt{CosmoMC}~\cite{Lewis:2002ah}, equipped with a convergence diagnostic based on the Gelman and Rubin statistic~\cite{Gelman:1992zz}, implementing an efficient sampling of the posterior distribution that makes use of the fast/slow parameter decorrelations \cite{Lewis:2013hha}. \texttt{CosmoMC} includes the support for the 2018 Planck data release~\cite{Aghanim:2019ame} (see \url{http://cosmologist.info/cosmomc/}).

\begin{table}
\begin{center}
\begin{tabular}{c|c}
Parameter                    & Prior\\
\hline 
$\Omega_{b} h^2$             & $[0.005,0.1]$\\
$\Omega_{c} h^2$             & $[0.001,0.99]$\\
$\tau$                       & $[0.01,0.8]$\\
$n_s$                        & $[0.8,1.2]$\\
$\log[10^{10}A_{s}]$         & $[1.6,3.9]$\\
$100\theta_{MC}$             & $[0.5,10]$\\ 
$M$                        & $[0.5,1]$\\ 
$\Omega_k$                        & $[-0.3,0.3]$\\
\end{tabular}
\end{center}
\caption{Flat priors adopted for the cosmological parameters. 
}
\label{tab:priors}
\end{table}

\section{Cosmological Constraints} \label{sec:results} 

Cosmological parameter constraints are summarized in Table~\ref{tab:novevflat} and \ref{tab:novevcurv} for the original 
VM case with no high redshift cosmological constant, i.e.\ no 
vacuum expectation value ``noVEV''. Table~\ref{tab:novevcurv} 
includes spatial curvature $\omk$ as a fit parameter. 
The 68\% and 95\% marginalized confidence level parameter 
contours and 1D PDFs are shown in Fig.~\ref{fig:noVEVflat} 
and Fig.~\ref{fig:noVEVcurvature}, respectively. 
Analogously, the VM VEV model results are presented in  
Table~\ref{tab:vmvevflat} for the VM VEV flat case and in
Table~\ref{tab:vmvevcurv} for the VM VEV curvature case, 
with Fig.~\ref{fig:VEVflat} and Fig.~\ref{fig:VEVcurvature} 
showing the parameter contours and PDFs.

\subsection{General Results} 

The first result we note is that the values of $H_0$ obtained in 
the VM model are significantly higher than in \lcdm, with values 
of $H_0\approx 73-74$ readily reached. Even though VM noVEV has 
the same number of parameters as \lcdm, the uncertainty on the 
$H_0$ determination from CMB alone is considerably higher: the 
equivalent CMB TTTEEE only constraint in \lcdm\ is $H_0=67.27\pm0.60$. 

A similar trend in the size of the $H_0$ uncertainty exists within $w$CDM (with one extra parameter), where the uncertainty nearly 
fills the priors. Thus \lcdm\ is a special case regarding the level of tension 
in $H_0$ for CMB data alone. If we include both BAO and SN data, 
then the uncertainty recedes to 0.66 (for VM noVEV flat), compared to 0.43 for \lcdm, 
and the mean value $H_0=74.21$ is quite consistent with R19, even 
though we did not use a R19 prior, while it is $H_0=67.74$ for \lcdm.

The five standard cosmological fit parameters are basically the 
same for VM noVEV and \lcdm, but derived parameters such as $\om$ and  
$\sigma_8$, in addition to $H_0$, can be quite different. Therefore 
it is important to examine the overall fit to the data, not just 
look at a single parameter. We compare the best fit $\chi^2$ values of each 
combination of data used, between VM noVEV flat and \lcdm\ (where the 
number of parameters are equal), between VM noVEV flat and VM noVEV 
curvature (with one extra parameter, $\omk$), and between VM noVEV 
curvature and \lcdm$+\omk$ (with the same number of parameters). We have also checked 
individual contributions to the $\chi^2$, 
e.g.\ from low $\ell$ CMB data (which agrees 
well with the respective \lcdm\ $\chi^2$ values). 

\begin{figure*}
\includegraphics[width=0.7\textwidth]{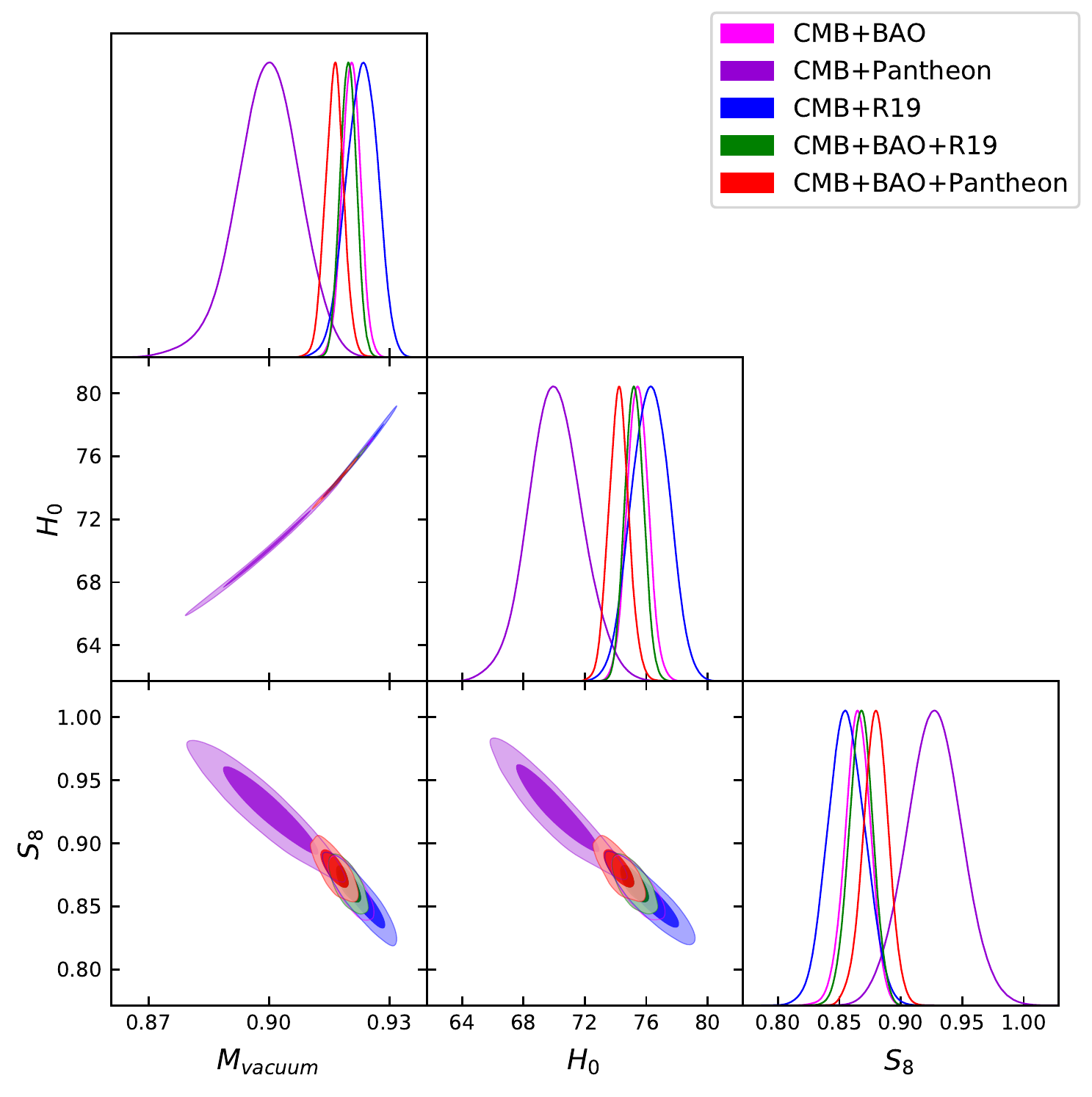}
\caption{68\% and 95\% CL constraints on the original VM case 
and spatial flatness: VM noVEV flat. }
\label{fig:noVEVflat}
\end{figure*}

\begin{figure*}
\includegraphics[width=0.7\textwidth]{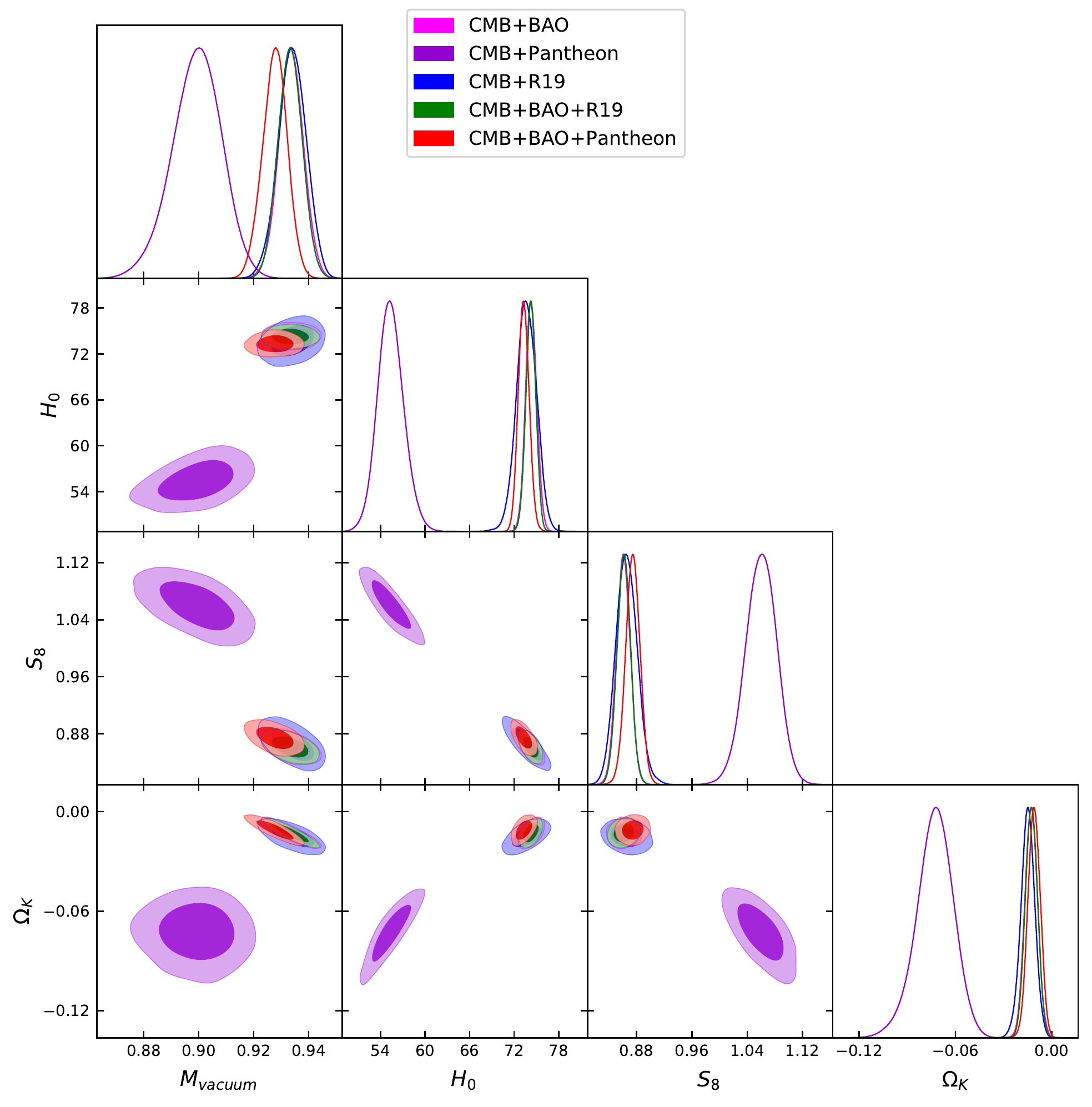}
\caption{68\% and 95\% CL constraints on the original VM case 
allowing 
spatial curvature to vary, i.e.\ VM noVEV curvature. }
\label{fig:noVEVcurvature}
\end{figure*}

Relative to \lcdm, the VM noVEV flat model (Table~\ref{tab:novevflat}  and 
Fig.~\ref{fig:noVEVflat}) has moderate improvements 
in $\chi^2$ for CMB data, a  strong improvement for CMB+R19, 
but much worse fits for CMB with BAO or SN. When both  models allow 
for curvature (Table~\ref{tab:novevcurv} 
and  
Fig.~\ref{fig:noVEVcurvature}), the CMB only fit becomes slightly worse, and the 
fits with BAO or SN improve significantly, but not enough to overtake 
\lcdm$+\omk$. And the combination CMB+BAO+SN shows 
significant tension, as we discuss in Sec.~\ref{sec:peak2}. Thus focusing only on $H_0$ gives a very biased view 
of the usefulness of a cosmological model. Within the 
VM noVEV models, the addition of curvature has a moderate improvement 
relative to the VM noVEV flat case for the CMB only fit, a strong to  very strong   
effect on the CMB with BAO or R19 or SN sets, and a significant effect on CMB+BAO+SN ($\Delta\chi^2=-10$ for one 
extra parameter). In this last combination the preference for a 
closed universe is $2.9\sigma$, but  again this model is a worse 
fit than \lcdm$+\omk$ -- which is consistent with flatness -- by $\Delta\chi^2=85$. The CMB+SN case prefers a quite distinct part of the posterior, and so 
while its fit is reasonable, the combination with BAO is 
emphatically not. The fit with CMB+BAO+R19 has relative  $\Delta\chi^2=-83$ with respect to CMB+BAO+SN when each is compared 
to the corresponding \lcdm+$\Omega_k$ case. Thus the VM noVEV model, whether flat or with curvature, though it does naturally give $H_0$ consistent with R19, cannot simultaneously 
satisfy CMB, BAO, and SN data.

\begin{table*}[tb]
\caption{ 68\% CL constraints on the cosmological parameters for the different dataset combinations explored in this work. This is for the original VM case 
and spatial flatness: VM noVEV flat. $\Delta\chi^2_{\rm bf}$ (best fit) is relative to the corresponding data best fits within \lcdm.
} 
\label{tab:novevflat} 
\begin{center}
\resizebox{\textwidth}{!}{  
\begin{tabular}{ c |c c c c c c c} 
  \hline
 \hline
 Parameters & CMB & CMB+lensing & CMB+BAO & CMB+Pantheon & CMB+R19  & CMB+BAO+Pantheon & CMB+BAO+R19 \\ 

 \hline
  $\Omega_b h^2$ & $0.02238\pm0.00014$ & $0.02242\pm0.00013$ & $0.02218\pm0.00012$ & $0.02201\pm0.00013$ & $0.02221\pm0.00012$ & $0.02213\pm0.00012$ & $0.02217\pm0.00012$ \\
  $100\theta_{MC}$ & $1.04091\pm0.00030$ & $1.04097\pm0.00029$ & $1.04060\pm0.00029$ & $1.04033\pm0.00031$ & $1.04063\pm0.00029$ & $1.04053\pm0.00029$ & $1.04060\pm0.00029$ \\
  $\tau$ & $0.0524\pm0.0078$ & $0.0510\pm0.0078$ & $0.0458^{+0.0083}_{-0.0067}$ & $0.039^{+0.010}_{-0.007}$ & $0.0469\pm0.0075$ & $0.0449^{+0.0079}_{-0.0065}$ & $0.0456^{+0.0083}_{-0.0068}$ \\
  $M$ & $0.9363^{+0.0055}_{-0.0044}$ & $0.9406\pm0.0034$ & $0.9205\pm0.0023$ & $0.8996^{+0.0081}_{-0.0073}$ & $0.9230^{+0.0042}_{-0.0036}$ & $0.9163\pm0.0023$ & $0.9198\pm0.0020$ \\
  ${\rm{ln}}(10^{10}A_s)$ & $3.041\pm0.016$ & $3.036\pm0.015$ & $3.035^{+0.017}_{-0.014}$ & $3.027^{+0.020}_{-0.014}$ & $3.036\pm0.016$ & $3.035^{+0.017}_{-0.014}$ & $3.035^{+0.017}_{-0.015}$ \\
  $n_s$ & $0.9643\pm0.0039$ & $0.9663\pm0.0036$ & $0.9572\pm0.0031$ & $0.9511\pm0.0036$ & $0.9585\pm0.0033$ & $0.9560\pm0.0031$ & $0.9571\pm0.0031$ \\
   \hline
  $H_0 {\rm[km/s/Mpc]}$ & $81.1\pm2.1$ & $82.9\pm1.5$ & $75.44\pm0.69$ & $70.1\pm1.8$ & $76.3\pm1.2$ & $74.21\pm0.66$ & $75.22\pm0.60$ \\
  $\sigma_8$ & $0.9440\pm0.0077$ & $0.9392\pm0.0067$ & $0.9456^{+0.0082}_{-0.0070}$ & $0.9419^{+0.0098}_{-0.0069}$ & $0.9457\pm0.0075$ & $0.9461^{+0.0080}_{-0.0068}$ & $0.9457^{+0.0082}_{-0.0073}$ \\
  $S_8$ & $0.805\pm0.022$ & $0.783\pm0.014$ & $0.865\pm0.010$ & $0.927\pm0.023$ & $0.856\pm0.015$ & $0.880\pm0.010$ & $0.8675\pm0.0098$ \\
  $\Omega_m$ & $0.218^{+0.010}_{-0.012}$ & $0.2085\pm0.0076$ & $0.2510\pm0.0046$ & $0.291\pm0.015$ & $0.2458^{+0.0074}_{-0.0084}$ & $0.2593\pm0.0046$ & $0.2525\pm0.0040$ \\
  \hline
  $\bar{\chi^2_{\rm bf}}$ & $2767.74$ & $2776.23$ & $2806.22$ & $3874.13$ & $2777.04$ & $3910.01$ & $2808.34$ \\
$\Delta \bar{\chi^2_{\rm bf}}$ & $-4.91$ & $-5.81$ & $+26.51$ & $+66.63$ & $-14.80$ & $+95.83$ & $+11.29$\\
   
 \hline
  \hline
\end{tabular}
}
\end{center}
\label{table}
\end{table*}

\begin{table*}[tb]
\caption{As Table~\ref{tab:novevflat} but also allowing 
spatial curvature to vary, i.e.\ VM noVEV curvature.} 
\label{tab:novevcurv} 
\begin{center}
\resizebox{\textwidth}{!}{  
\begin{tabular}{ c |c c c c c c c} 
  \hline
 \hline
 Parameters & CMB & CMB+lensing & CMB+BAO & CMB+Pantheon & CMB+R19  & CMB+BAO+Pantheon & CMB+BAO+R19 \\ 

 \hline
  $\Omega_b h^2$ & $0.02263\pm0.00017$ & $0.02250\pm0.00016$ & $0.02251\pm0.00016$ & $0.02275\pm0.00016$ & $0.02254\pm0.00016$ & $0.02239\pm0.00015$ & $0.02250\pm0.00016$ \\
  $100\theta_{MC}$ & $1.04119\pm0.00032$ & $1.04107\pm0.00032$ & $1.04105\pm0.00033$ & $1.04130\pm0.00033$ & $1.04111\pm0.00031$ & $1.04092\pm0.00031$ & $1.04105\pm0.00032$ \\
  $\tau$ & $0.0486^{+0.0084}_{-0.0075}$ & $0.0498\pm0.0083$ & $0.0511\pm0.0075$ & $0.0391^{+0.0011}_{-0.0007}$ & $0.0512\pm0.0079$ & $0.0483^{+0.0079}_{-0.0067}$ & $0.0513\pm0.0075$ \\
  $\Omega_k$ & $-0.030^{+0.016}_{-0.011}$ & $-0.0044^{+0.0064}_{-0.0052}$ & $-0.0128\pm0.0039$ & $-0.073^{+0.012}_{-0.010}$ & $-0.0149\pm0.0044$ & $-0.0110\pm0.0038$ & $-0.0127\pm0.0037$ \\
  $M$ & $0.927^{+0.010}_{-0.007}$ & $0.9416\pm0.0037$ & $0.9336\pm0.0044$ & $0.8993^{+0.0094}_{-0.0083}$ & $0.9340\pm0.0050$ & $0.9277\pm0.0044$ & $0.9333\pm0.0043$ \\
  ${\rm{ln}}(10^{10}A_s)$ & $3.028^{+0.018}_{-0.016}$ & $3.031\pm0.018$ & $3.035\pm0.015$ & $3.006^{+0.023}_{-0.014}$ & $3.034\pm0.016$ & $3.033^{+0.016}_{-0.014}$ & $3.036\pm0.016$ \\
  $n_s$ & $0.9711\pm0.0047$ & $0.9687\pm0.0047$ & $0.9684\pm0.0045$ & $0.9743\pm0.0046$ & $0.9692\pm0.0045$ & $0.9648\pm0.0043$ & $0.9680\pm0.0044$ \\
   \hline
  $H_0 {\rm[km/s/Mpc]}$ & $67.8^{+4.9}_{-5.4}$ & $81.0\pm3.1$ & $74.30\pm0.74$ & $55.4^{+1.6}_{-1.8}$ & $73.7\pm1.3$ & $73.30\pm0.72$ & $74.25\pm0.67$ \\
  $\sigma_8$ & $0.903^{+0.022}_{-0.018}$ & $0.931\pm0.013$ & $0.9259\pm0.0091$ & $0.848^{+0.015}_{-0.014}$ & $0.9226\pm0.0098$ & $0.9284\pm0.0094$ & $0.9262\pm0.0093$ \\
  $S_8$ & $0.925\pm0.053$ & $0.795\pm0.022$ & $0.8609\pm0.0099$ & $1.060\pm0.022$ & $0.866\pm0.015$ & $0.875\pm0.010$ & $0.8616\pm0.0097$ \\
  $\Omega_m$ & $0.318^{+0.040}_{-0.055}$ & $0.219^{+0.015}_{-0.018}$ & $0.2594\pm0.0052$ & $0.469\pm0.030$ & $0.264^{+0.009}_{-0.010}$ & $0.2663\pm0.0053$ & $0.2597\pm0.0047$ \\
  \hline
  $\bar{\chi^2_{\rm bf}}$ & $2762.63$ & $2777.66$ & $2795.27$ & $3811.16$ & $2763.16$ & $3899.96$ & $2795.92$ \\
$\Delta \bar{\chi^2_{\rm bf}}$ & $+2.54$ & $+1.60$ & $+18.29$ & $+3.94$ & $-19.84$ & $+85.48$ &$+2.45$\\
  
 \hline
  \hline
\end{tabular}
}
\end{center}
\label{table}
\end{table*}

\begin{figure*}
\includegraphics[width=0.7\textwidth]{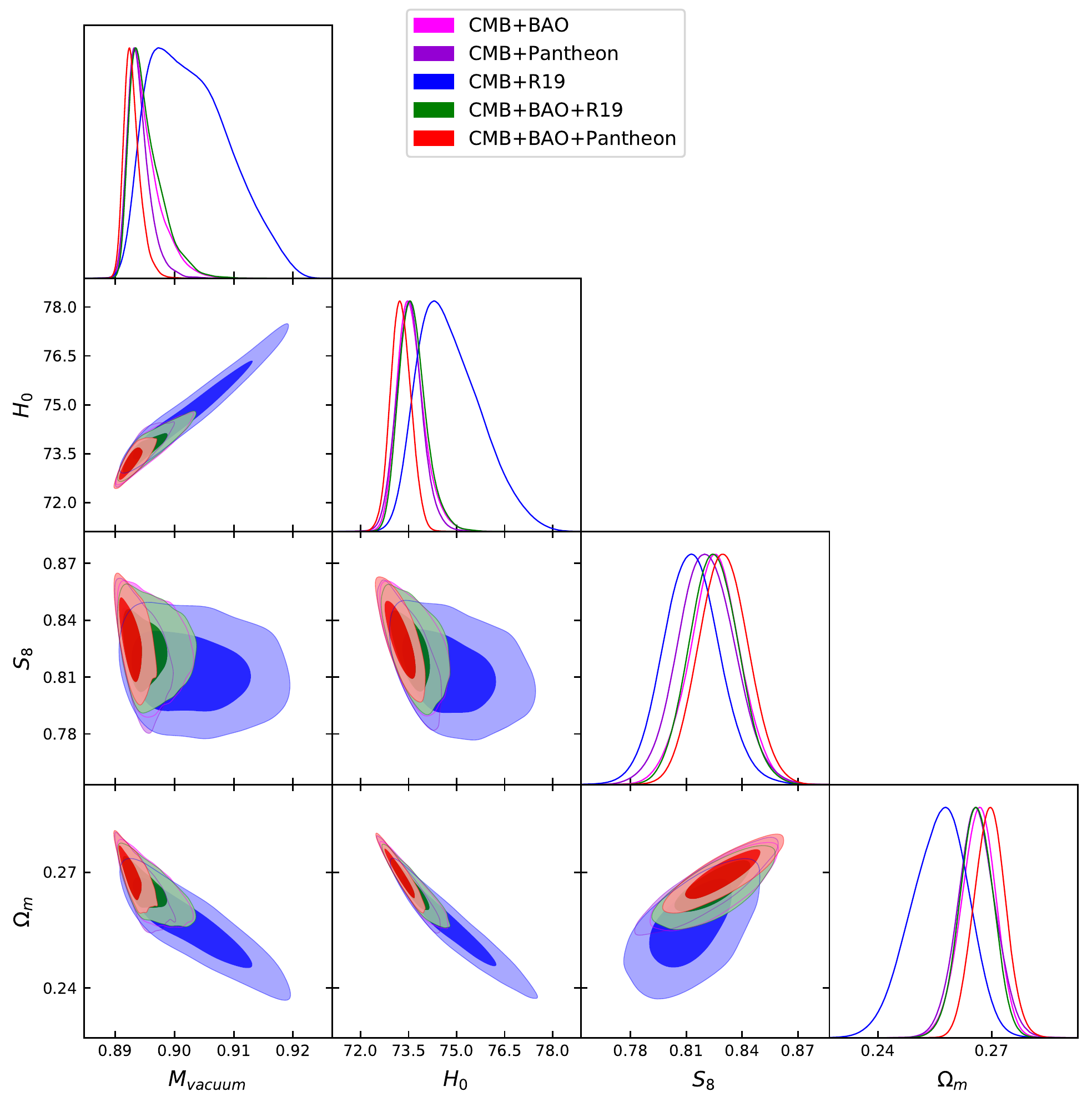}
\caption{68\% and 95\% CL constraints on the VM VEV model, i.e.\ VM VEV flat. }
\label{fig:VEVflat}
\end{figure*}

\begin{figure*}
\includegraphics[width=0.7\textwidth]{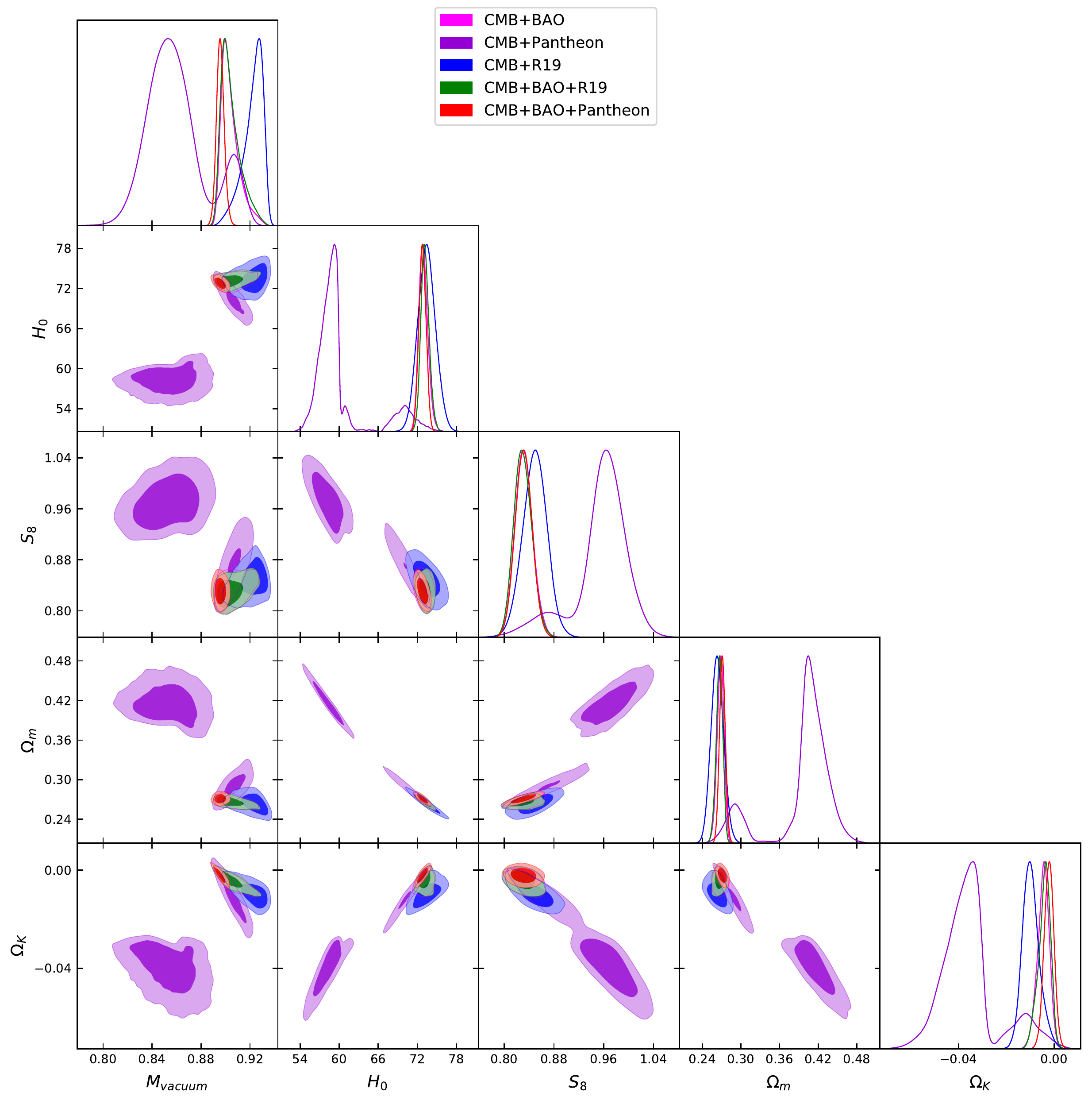}
\caption{68\% and 95\% CL constraints on the VM VEV case 
allowing 
spatial curvature to vary, i.e.\ VM VEV curvature.
}
\label{fig:VEVcurvature}
\end{figure*}

Moving to the VM VEV model, we present the results in  
Table~\ref{tab:vmvevflat} 
and Fig.~\ref{fig:VEVflat} for the VM VEV flat case and 
Table~\ref{tab:vmvevcurv} 
and Fig.~\ref{fig:VEVcurvature} for the VM VEV curvature case. 
The addition of one parameter for the 
VEV (i.e.\ both $M$ and $\Omega_c h^2$ or effectively $\om$ free) 
has an insignificant effect relative to VM noVEV
for CMB only, a strong effect for CMB 
with BAO or SN, and a highly significant effect for the combination 
CMB+BAO+SN ($\Delta\chi^2=-53$ for one extra parameter). The 
VM VEV model brings $H_0\approx73$, $\om\approx0.27$, and $S_8\approx 0.83$, whether flat or with curvature. 
Relative to \lcdm, the VM VEV flat model has $\Delta\chi^2=+43$ with one extra parameter for CMB+BAO+SN. 
Note that \cite{Rubin:2008wq} also showed the tension of fitting 
CMB+BAO+SN simultaneously (for the data at that time), where in their Fig.~6 the SN contour  
curves away from the others as one reduces the VM VEV case to  
VM noVEV, so the SN data play a particularly important discriminating role. 
While the fit with CMB+BAO+R19 actually has an improved fit 
relative to \lcdm, this comes at the price of neglecting the 
SN constraints; we discuss this further in Sec.~\ref{sec:peak2}. 

Within the VM VEV model, the addition of curvature has a modest  
effect, except a strong improvement for CMB+SN (simply  due to curvature washing out a great part of SN's probative power suppressing one of posterior peaks discussed in Sec.~\ref{sec:peak2}). However, CMB+BAO+SN (and CMB+BAO+R19) 
shows insignificant change. CMB, CMB+BAO, and CMB+SN all prefer 
a closed universe but this reduces to a $\sim1\sigma$ effect for 
CMB+BAO+SN. For VM VEV curvature relative to 
\lcdm$+\omk$, $\Delta\chi^2=+40$. Removing the SN
information gives results relative to \lcdm$+\omk$ of 
$\Delta\chi^2=+10$ for CMB+BAO and $-6$ for CMB+BAO+R19. Note that adding 
R19 does not particularly improve the VM VEV fit -- for one extra 
data point $\Delta\chi^2=+0.5$ -- but the apparent gain 
is caused by the \lcdm$+\omk$ fit getting worse by 16. 

Thus, while the VM model can relieve the tension between CMB data and 
the R19 local Cepheid distance ladder measurement of $H_0$, it is not as good 
a fit to cosmology across a wider combination of data sets including 
BAO or SN or both. The need for a wide range of probes, 
including growth ones, will be the main theme of
Sec.~\ref{sec:conjoin}, but first we examine  
apparent sources of tension between CMB+BAO and CMB+SN, and 
summarize the constraints on $H_0$ and $\omk$.

\begin{table*}[tb]
\caption{As Table~\ref{tab:novevflat} but for the 
VM VEV model, i.e.\ VM VEV flat.} 
\label{tab:vmvevflat} 
\begin{center}
\resizebox{\textwidth}{!}{  
\begin{tabular}{ c |c c c c c c c} 
  \hline
 \hline
 Parameters & CMB & CMB+lensing & CMB+BAO & CMB+Pantheon & CMB+R19  & CMB+BAO+Pantheon & CMB+BAO+R19 \\  

 \hline
  $\Omega_b h^2$ & $0.02238\pm0.00015$ & $0.02242\pm0.00015$ & $0.02229\pm0.00014$ & $0.02233\pm0.00015$ & $0.02236\pm0.00015$ & $0.02228\pm0.00014$ & $0.02230\pm0.00014$ \\
  $\Omega_c h^2$ & $0.1200\pm0.0013$ & $0.1194\pm0.0012$ & $0.1213\pm0.0012$ & $0.1208\pm0.0014$ & $0.1203\pm0.0014$ & $0.1217\pm0.0012$ & $0.1212\pm0.0011$ \\
  $100\theta_{MC}$ & $1.04092\pm0.00031$ & $1.04098\pm0.00030$ & $1.04079\pm0.00030$ & $1.04086\pm0.00031$ & $1.04090\pm0.00032$ & $1.04077\pm0.00030$  & $1.04080\pm0.00031$ \\
  $\tau$ & $0.0541\pm0.0078$ & $0.0529\pm0.0076$ & $0.0527\pm0.0077$ & $0.0529\pm0.0077$ & $0.0537\pm0.0079$ & $0.0524\pm0.0078$ & $0.0530\pm0.0077$ \\
  $M$ & $0.914^{+0.021}_{-0.009}$ & $0.920^{+0.017}_{-0.007}$ & $0.8950^{+0.0013}_{-0.0033}$ & $0.8940^{+0.0012}_{-0.0022}$ & $0.9028^{+0.0046}_{-0.0085}$ & $0.8929^{+0.0010}_{-0.0016}$ & $0.8953^{+0.0014}_{-0.0034}$ \\
  ${\rm{ln}}(10^{10}A_s)$ & $3.044\pm0.016$ & $3.039\pm0.015$ & $3.044\pm0.016$ & $3.043\pm0.016$ & $3.044\pm0.016$ & $3.044\pm0.016$ & $3.045\pm0.016$ \\
  $n_s$ & $0.9653\pm0.0044$ & $0.9666\pm0.0040$ & $0.9620\pm0.0041$ & $0.9632\pm0.0025$ & $0.9644\pm0.0044$ & $0.9612\pm0.0040$ & $0.9623\pm0.0038$ \\
   \hline
  $H_0 {\rm[km/s/Mpc]}$ & $76.7^{+3.9}_{-2.6}$ & $78.0^{+3.2}_{-1.9}$ & $73.58^{+0.33}_{-0.49}$ & $73.53^{+0.37}_{-0.42}$ & $74.8^{+0.7}_{-1.2}$ & $73.26\pm0.32$ & $73.63^{+0.33}_{-0.48}$ \\
  $\sigma_8$ & $0.895^{+0.016}_{-0.026}$ & $0.900^{+0.024}_{-0.019}$ & $0.876\pm0.010$ & $0.872\pm0.010$ & $0.880^{+0.012}_{-0.016}$ & $0.8756\pm0.0091$ & $0.8760^{+0.0093}_{-0.0099}$ \\
  $S_8$ & $0.805\pm0.016$ & $0.796^{+0.013}_{-0.015}$ & $0.825\pm0.014$ & $0.821\pm0.015$ & $0.813\pm0.015$ & $0.830\pm0.013$ & $0.825\pm0.013$ \\
  $\Omega_m$ & $0.243^{+0.017}_{-0.025}$ & $0.235^{+0.011}_{-0.020}$ & $0.2664^{+0.0048}_{-0.0043}$ & $0.2661\pm0.0050$ & $0.2561^{+0.0081}_{-0.0068}$ & $0.2695\pm0.0041$ & $0.2660\pm0.0044$ \\
  \hline
  $\bar{\chi^2_{\rm bf}}$ & $2769.74$ & $2778.93$ & $2790.75$ & $3840.55$ & $2772.09$ & $3857.21$ & $2789.76$ \\
$\Delta \bar{\chi^2_{\rm bf}}$ & $-2.91$ & $-3.11$ & $+11.04$ & $+33.05$ & $-19.75$ & $+43.03$&$-7.29$ \\
  
 \hline
  \hline
\end{tabular}
}
\end{center}
\label{table}
\end{table*}

\begin{table*}[tb]
\caption{As Table~\ref{tab:vmvevflat} but also allowing spatial curvature to vary, i.e.\ VM VEV curvature.  
Italicized entries involving Pantheon data serve as a reminder that 
the parameter values are the mean over both peaks in the posterior; a prior such as $H_0\gtrsim60$ would select the peak that is much more consistent with the other data though it does not much affect the total $\chi^2$ values.  
} 
\label{tab:vmvevcurv} 
\begin{center}
\resizebox{\textwidth}{!}{  
\begin{tabular}{ c |c c c c c c c} 
  \hline
 \hline
 Parameters & CMB & CMB+lensing & CMB+BAO & CMB+Pantheon & CMB+R19  & CMB+BAO+Pantheon & CMB+BAO+R19 \\ 

 \hline
  $\Omega_b h^2$ & $0.02262\pm0.00016$ & $0.02248\pm0.00016$ & $0.02243\pm0.00015$ & $0.02261\pm0.00017$ & $0.02251\pm0.00016$ & $0.02235\pm0.00015$ & $0.02243\pm0.00016$ \\
  $\Omega_c h^2$ & $0.1180\pm0.0015$ & $0.1187\pm0.0015$ & $0.1196\pm0.0014$ & $0.1181^{+0.0015}_{-0.0017}$ & $0.1188\pm0.0014$ & $0.1207\pm0.0014$ & $0.1197\pm0.0014$ \\
  $100\theta_{MC}$ & $1.04119\pm0.00033$ & $1.04105\pm0.00032$ & $1.04098\pm0.00031$ & $1.04118\pm0.00033$ & $1.04104^{+0.00035}_{-0.00030}$ & $1.04086\pm0.00032$ & $1.04097\pm0.00032$ \\
  $\tau$ & $0.0482\pm0.0080$ & $0.0506\pm0.0080$ & $0.0537\pm0.0075$ & $0.0480^{+0.0082}_{-0.0071}$ & $0.0519\pm0.0076$ & $0.0528\pm0.0077$ & $0.0537\pm0.0079$ \\
  $\Omega_k$ & $-0.040\pm0.014$ & $-0.0034\pm0.0043$ & $-0.0045^{+0.0027}_{-0.0021}$ & $\mathit{-0.035^{+0.005}_{-0.011}}$ & $-0.0096^{+0.0029}_{-0.0039}$ & $-0.0021\pm0.0020$ & $-0.0042^{+0.0030}_{-0.0020}$ \\
  $M$ & $0.826^{+0.034}_{-0.072}$ & $0.919^{+0.015}_{-0.012}$ & $0.9040^{+0.0038}_{-0.0089}$ & $\mathit{0.861^{+0.013}_{-0.029}}$ & $0.922^{+0.010}_{-0.004}$ & $0.8958^{+0.0029}_{-0.0032}$ & $0.9045^{+0.0043}_{-0.0098}$ \\
  ${\rm{ln}}(10^{10}A_s)$ & $3.027\pm0.017$ & $3.033\pm0.016$ & $3.042\pm0.015$ & $3.027^{+0.017}_{-0.015}$ & $3.037\pm0.016$ & $3.043\pm0.016$ & $3.042\pm0.016$ \\
  $n_s$ & $0.9710\pm0.0046$ & $0.9681\pm0.0046$ & $0.9664\pm0.0046$ & $0.9707\pm0.0047$ & $0.9683\pm0.0045$ & $0.9636\pm0.0045$ & $0.9661\pm0.0045$ \\
   \hline
  $H_0 {\rm[km/s/Mpc]}$ & $58.2^{+1.3}_{-4.8}$ & $76.2^{+2.3}_{-4.2}$ & $72.96^{+0.59}_{-0.76}$ & $\mathit{60.2^{+0.2}_{-3.6}}$ & $73.4\pm1.3$ & $72.80\pm0.55$ & $73.19^{+0.56}_{-0.69}$ \\
  $\sigma_8$ & $0.818^{+0.013}_{-0.043}$ & $0.894^{+0.018}_{-0.028}$ & $0.880^{+0.010}_{-0.015}$ & $\mathit{0.835^{+0.009}_{-0.029}}$ & $0.906^{+0.019}_{-0.013}$ & $0.8743\pm0.0093$ & $0.881^{+0.010}_{-0.016}$ \\
  $S_8$ & $0.969^{+0.041}_{-0.031}$ & $0.807^{+0.022}_{-0.020}$ & $0.832\pm0.014$ & $\mathit{0.955^{+0.045}_{-0.020}}$ & $0.849\pm0.019$ & $0.831\pm0.013$ & $0.830^{+0.014}_{-0.015}$ \\
  $\Omega_m$ & $0.426^{+0.064}_{-0.031}$ & $0.245^{+0.025}_{-0.017}$ & $0.2681^{+0.0057}_{-0.0049}$ & $\mathit{ 0.395^{+0.043}_{-0.005}}$ & $0.2635\pm0.0098$ & $0.2712\pm0.0045$ & $0.2665^{+0.0053}_{-0.0047}$ \\
  \hline
  $\bar{\chi^2_{\rm bf}}$ & $2762.79$ & $2778.62$ & $2787.06$ & $\mathit{ 3801.45}$ & $2764.02$ & $3854.88$ & $2787.57$ \\
$\Delta \bar{\chi^2_{\rm bf}}$ & $+2.70$ & $+2.56$ & $+10.08$ & $\mathit{-5.77}$ & $-18.98$ & $+40.40$&$-5.90$ \\
  
 \hline
  \hline
\end{tabular}
}
\end{center}
\label{table}
\end{table*}

\subsection{BAO and SN Constraining Power} \label{sec:peak2} 

In Fig.~\ref{fig:noVEVcurvature} we see a CMB+SN confidence contour 
disjoint from CMB+BAO, and in Fig.~\ref{fig:VEVcurvature} we 
see in addition multipeaked 1D PDFs. It 
is worthwhile understanding their origin and effects. First, we 
emphasize that all chains are well converged, with $R-1<0.02$, so these are real features of the posteriors. 

The issue can be traced back to the CMB constraints, so it is not 
due to SN. The CMB carries cosmological information in three main 
characteristics: the geometric degeneracy to the last scattering 
surface (location of the acoustic peaks), the acoustic peak structure 
(influencing the baryon and cold dark matter constraints), and the 
integrated Sachs-Wolfe effect. In the VM noVEV model with curvature, 
and the VM VEV models both with and without curvature, 
there is sufficient freedom to prolong the matter domination era 
by raising $\om$ (compensating with $\omk$ to preserve the geometric 
degeneracy) and thus suppress the ISW effect. This brings the lower 
$\ell$ multipoles into better agreement with CMB data. The intersection of the ISW suppression 
relation and geometric degeneracy in the $M$--$\om$ plane gives 
an opportunity for an extended CMB confidence contour, or second peak in the posterior. 

For the VM noVEV model with curvature, the SN likelihood enhances 
the low ISW part of the confidence region because this also has 
larger $\om$ to offset the curvature for the SN distances. That is, 
the addition of the SN data chooses the far end of the CMB confidence 
region continuously connecting the two regions seen from CMB+SN 
and CMB+BAO. For the VM VEV cases the ISW suppression is stronger 
for the parameter region corresponding to the upper limit of $M$ in 
Eq.~(\ref{eq:ombounds}), and this actually creates a second  
peak in the CMB posterior. Without the degeneracy due to curvature, 
the SN data is informative enough to nullify this second peak, just 
as BAO do, and so Fig.~\ref{fig:VEVflat} shows CMB+BAO and CMB+SN 
in substantial agreement. However, again SN leverage is weakened 
in the presence of curvature and it is not able to remove the 
strongest part of the second posterior region from the CMB, giving 
rise to a disjoint second contour 
in Fig.~\ref{fig:VEVcurvature}. The unusual parameter values for 
CMB+SN in this case is due to the disjoint part of the 
CMB+SN posterior -- the $H_0\gtrsim60$ disjoint part of the CMB+SN 
posterior has reasonable consistency with BAO data, e.g.\ with 
$H_0\approx72$, $\om\approx0.27$. 
To highlight this we have 
italicized in Table~\ref{tab:vmvevcurv} the entries involving Pantheon data to serve as a reminder that 
the values are the mean over both peaks in the posterior. 
Removing the $H_0<60$ contour 
peak, though, does not change $\chi^2_{\rm bf}$  for CMB+BAO+SN.

\subsection{$H_0$ Results} \label{sec:h0} 

As we have seen, the VM model can accommodate a high (R19) value of $H_0$ while
being consistent with the CMB. Indeed the flat VM models have a better 
CMB $\chi^2$ than \lcdm, with the same number of parameters, while giving 
high $H_0$. Including curvature, the VM models do about as well as 
\lcdm$+\omk$ for CMB data, with greater uncertainties on $H_0$ but 
consistent with either low (Planck \lcdm) or high values (the VM VEV curvature case 
for CMB alone has a peak at $H_0\sim70$ between the low and high values, 
as well as its main peak at $H_0\sim55$). These high values occur 
naturally, without adding an external $H_0$ prior. 

\begin{figure*} 
\includegraphics[width=0.9\textwidth]{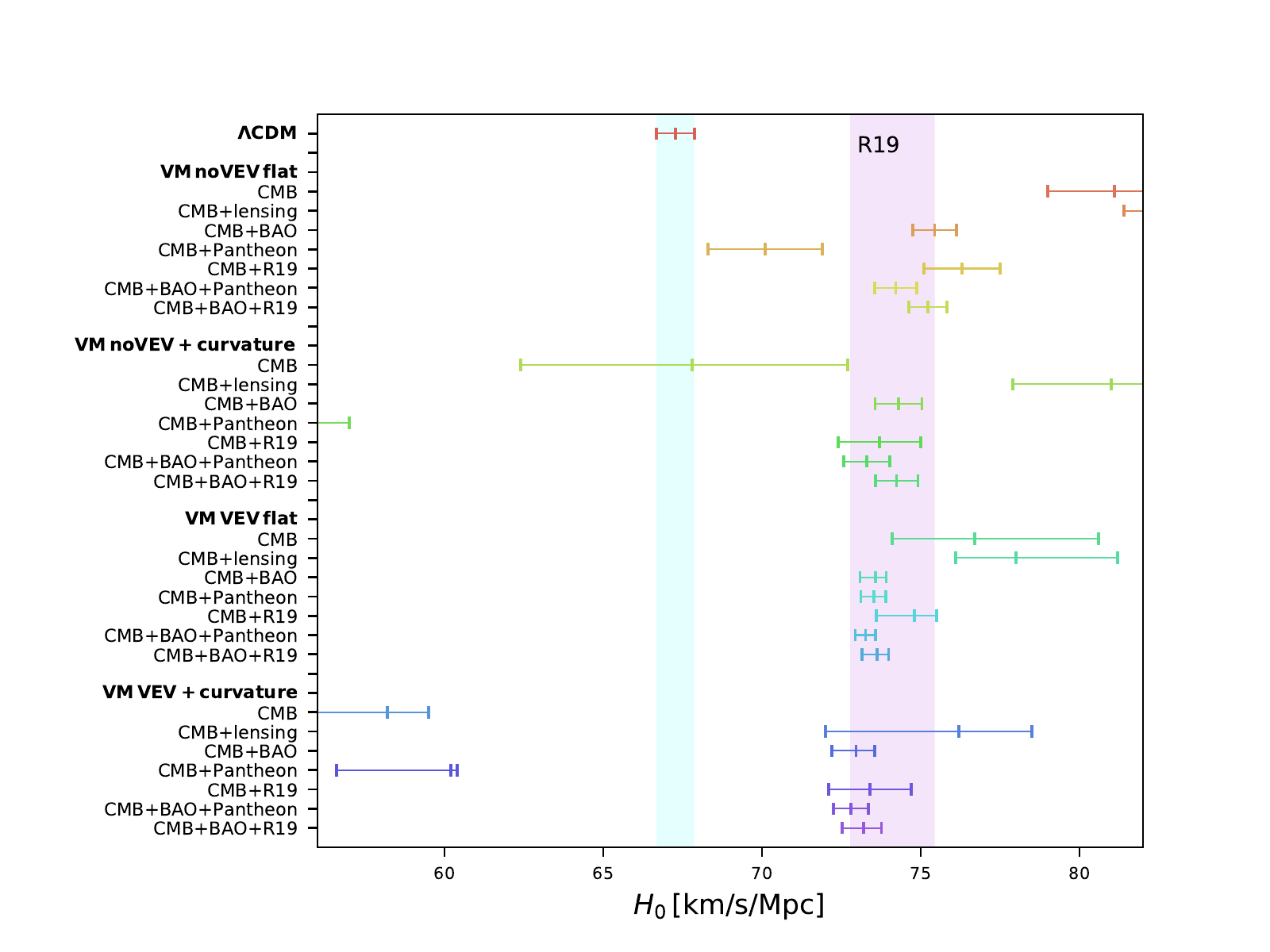} 
\caption{68\% CL constraints on $H_0$ for the different combinations of datasets and models explored here. The cyan region is the Planck $\Lambda$CDM estimate of the Hubble constant, while the violet region is the R19 measurement. 
}
\label{fig:H0}
\end{figure*}

Adding further data such as BAO and SN, or both, does not change the 
consistency with high $H_0$ (see the previous subsection for discussion 
of the CMB+SN data in the presence of curvature), although it can worsen 
(considerably) the overall goodness of fit. Figure~\ref{fig:H0} shows 
the $H_0$ values fit for all the data combinations, for all the VM 
model permutations. However, as we caution in Sec.~\ref{sec:conjoin}, 
using only the criterion of a fit to $H_0$ can be highly problematic.

\subsection{$\omk$ Results} \label{sec:omk} 

Evidence for nonzero spatial curvature, $\omk\ne0$, would have a profound 
impact on theories of the early universe and inflation. While we have seen 
that degeneracies with other parameters exist, and can increase uncertainties 
or easily allow large systematic shifts, the combination of many distinct 
observational probes can alleviate this. Furthermore, the VM model has the 
benefit of not increasing the number of parameters (and hence potential for 
further degeneracy) beyond \lcdm\ (for VM noVEV; VM VEV has a single added 
parameter). 

Earlier work, e.g.\ \cite{Handley:2019tkm,DiValentino:2019qzk,Efstathiou:2020wem,Liao:2020zko,DiValentino:2020hov}, has 
discussed evidence for (or against) a closed universe, $\omk<0$. However, 
these have considered only one or two probes together. Here we have several 
cases where we use three, either CMB+BAO+SN or CMB+BAO+R19. We do find that 
$\omk$ tends to lie closer to zero when combining three probes than two, but 
the results still lie $\sim3\sigma$ away from flatness for VM noVEV and 
$\sim1\sigma$ away for the better fitting VM VEV. 

\begin{figure*}
\includegraphics[width=0.8\textwidth]{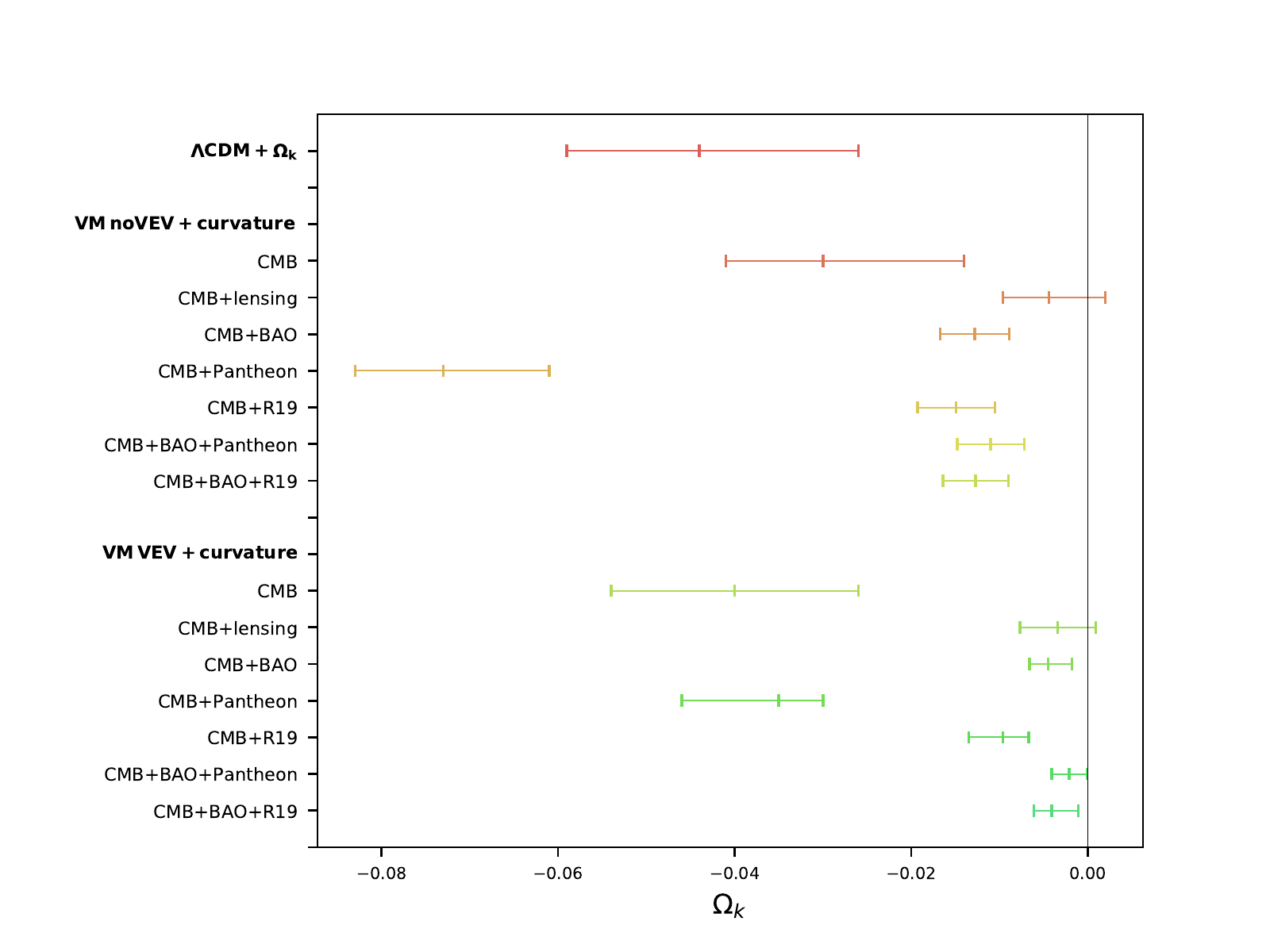}
\caption{68\% CL constraints on $\Omega_k$ for the different combinations 
of datasets and models explored here. 
}
\label{fig:omegak}
\end{figure*}

Figure~\ref{fig:omegak} shows 
the $\omk$ values fit for all the data combinations, for all the VM 
model permutations. Again, we caution in Sec.~\ref{sec:conjoin} that 
one should consider the full set of cosmological model parameters together, 
rather than only focus on one. (For example, the CMB only \lcdm$+\omk$ case 
that has an improvement of $\Delta\chi^2=-11$ over flat \lcdm\ also has 
$H_0\approx54$ and $\om\approx 0.48$; the VM VEV case is only slightly 
better, while the VM noVEV case is much closer to standard, both departing 
from flatness at $\sim2$--$3\sigma$.)

\section{Beyond $H_0$} \label{sec:conjoin} 

As we have seen, for some of the VM models, while $H_0$ is in good 
agreement with R19 local measurements, and the CMB only data is 
well fit, the parameters associated with the matter density and 
structure growth are pushed away from the standard \lcdm\ values. 
We explore the impact of this here, and show how the conjoined 
expansion-growth plots, e.g.\ as introduced in \cite{Linder:2016xer},  
can play an important role in assessing the overall fit of a 
model beyond just $H_0$. 

For example, focusing on the full data set combination CMB+BAO+SN, 
for \lcdm\ the matter density takes values around $\om=0.31$ 
while for the VM models it is $\om\approx0.27$. The 
structure growth parameters are more difficult to assess, since  
the perturbation theory for the VM theory after the transition 
where the scalar curvature $R$ freezes has not been calculated 
in detail (see \cite{Caldwell:2005xb} for a discussion of the difficulties). 
Thus, growth parameters such as $\sigma_8$ and $S_8$ should not 
be literally interpreted as values from the VM theory; however, 
they can be viewed as values for a dark energy theory with the 
same background expansion as VM (see Fig.~1 of 
\cite{DiValentino:2017rcr} for the sharp transition of $w(z)$ to a phantom behavior). That is how we will interpret 
$\sigma_8$ and $S_8$ for the remainder of this section. The 
Monte Carlo runs of the previous section are robust since we 
chose data sets -- CMB, BAO, and SN -- that are predominantly 
geometric measures and insensitive to the behavior of the dark 
energy perturbations after the transition, $z\lesssim1.5$. 

Examining $\sige$ in that light, we see a common characteristic 
of late time phantom transitions in dark energy: if they also 
seek to match the CMB, not just $H_0$, then they tend to have 
higher $\sige$ than \lcdm. The extended era of matter domination, 
i.e.\ decreased dark energy density at redshifts $z\approx0.5$--2, 
provides additional growth, increasing $\sigma_8$. For VM-like 
expansion, $\sige$ ranges from $\sim0.87$ for the VM VEV 
expansion models to $\sim 0.94$ for VM noVEV expansion models, 
quite a bit above $\sim0.81$ 
for \lcdm. Note that early time dark energy models that raise 
$H_0$ by adding early energy density and hence decreasing the 
sound horizon scale, also generally raise $\sige$ as well since 
they require a higher primordial curvature perturbation amplitude 
to offset the damping effect of the unclustered component. So 
both proposals for raising $H_0$ run into increased tension 
with values of $\sige$ found from structure constraints. 

However, many large scale structure surveys do not give  
$\sige$ per se, but rather a quantity close to 
$S_8\equiv \sige\,(\om/0.3)^{0.5}$. For example, this is  
the main degeneracy direction for many weak lensing shape 
measurements. This is also closely related to $\fsig(z=0)$, 
central to galaxy clustering redshift space distortions,  
where $f=\om(a)^{0.55}$ is an excellent approximation to  
the growth rate even for many non-\lcdm\ models 
\cite{2005PhRvD..72d3529L,2007APh....28..481L,wangs}. So for $S_8$, a redshift 
$z=0$ quantity, 
\be 
S_8=\fsig(z=0)\,\times \left(\frac{1}{0.3}\right)^{0.55}\, 
\left(\frac{\om}{0.3}\right)^{-0.05}\approx  1.94\,\fsig(z=0)\,.  
\ee 

Thus either weak lensing or redshift surveys give a quantity 
closer to $S_8$. As we have seen, many of the models that give  
a higher $H_0$ give a lower $\om$ if they also match CMB data.  
We see that for VM-like expansion models, $S_8$ tends to be 
$\sim0.83$ for the VM VEV cases, and $\sim  0.88$ for the VM 
noVEV cases, compared to $\sim0.82$ for \lcdm. Thus the 
VM VEV expansion cases give fairly good agreement with \lcdm\ 
on this structure parameter. Of course some probes such as 
galaxy clusters and the Sunyaev-Zel'dovich effect do measure 
$\sige$ more directly. We also must remember $S_8$ is simply 
the redshift $z=0$ value, and does not ensure that the growth 
history at $z=0.5$ or 1 is consistent with observational 
constraints. 

What we want is a compact method for assessing the value of 
$H_0$, the expansion history (as in CMB, BAO, and SN), and 
the growth history. This is often called a conjoined history 
diagram, highlighted in \cite{Linder:2016xer}. By plotting $\fsig(z)$ 
directly against $H(z)$, with redshift running along the 
evolutionary tracks, one can not only see distinctions in 
dark energy properties such as equation of state, but 
modifications of gravity, or phenomenological effects from 
``stuttered'' growth \cite{Linder:2016xer}. (Note that the
superdeceleration leading to stuttered growth is precisely 
what is necessary for an early time dark energy transition to 
raise the value of $H_0$ without disrupting agreement with 
CMB data.) 

\begin{figure}[tbp!] 
\includegraphics[width=0.47\columnwidth]{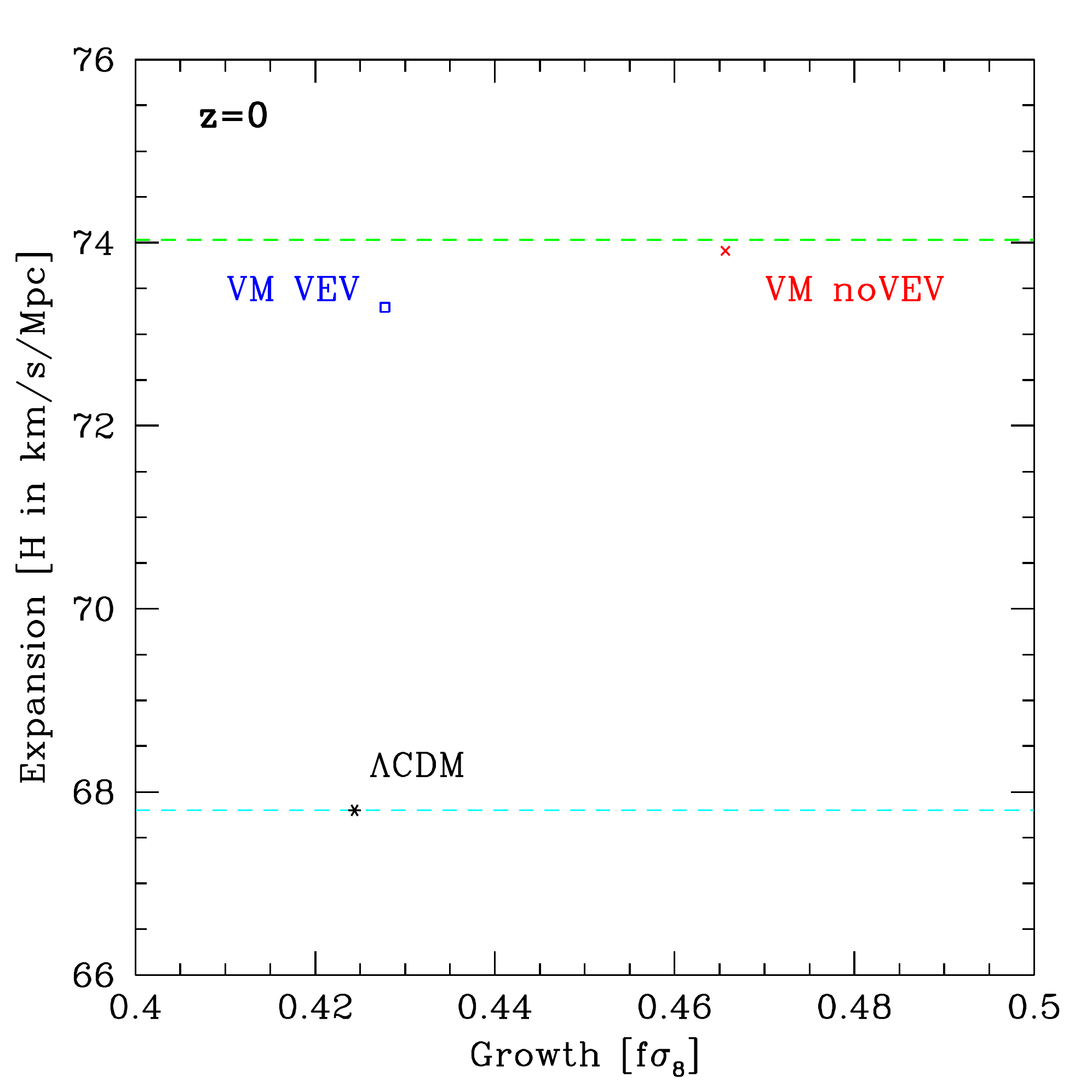}\ \ 
\includegraphics[width=0.47\columnwidth]{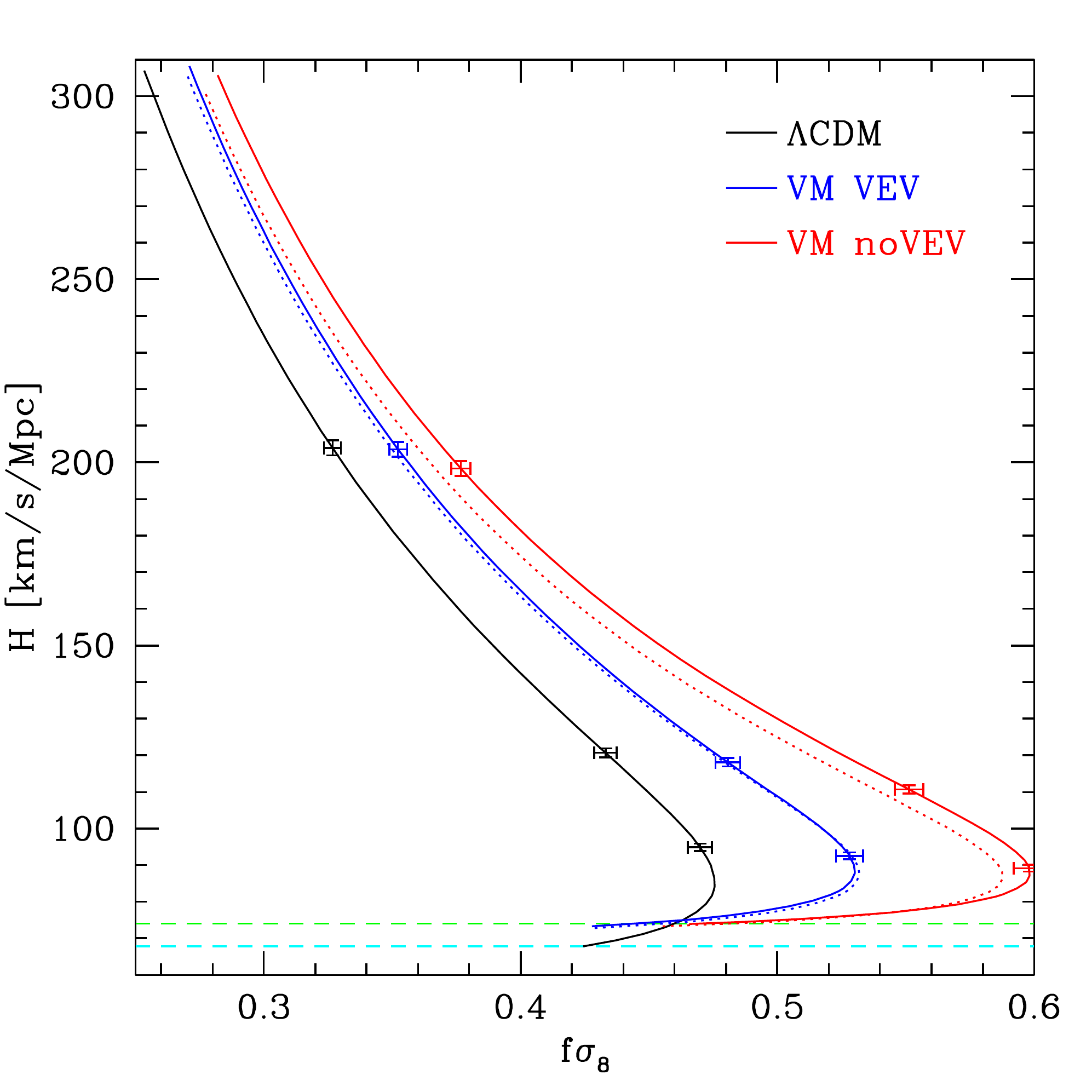}
\caption{
Expansion and growth histories are here plotted simultaneously. The left panel shows the situation looking only at $z=0$, hence $H_0$ and $\fsig(0)$ or effectively  $S_8$. Excellent agreement with the R19 value of $H_0$ (green dashed line),  as opposed to the \lcdm\ value (cyan dashed line) is obtained by the flat VEV models with parameters fit to  CMB+BAO+SN, and the VM VEV model agrees well on $S_8$ as well. However, the  right panel shows a very different situation if one looks beyond $H_0$ at the full conjoint  evolutionary track of the  various models, not just the $z=0$ (lower) endpoints shown in the left panel. Over the histories the VM 
models diverge considerably from \lcdm. Curves extend 
from $z=0$ at the bottom to $z=3$ at the top, and
the points with 
error bars show the trajectory location at $z=0.6$, 1, 
2, where the error bars mimic 
1\% constraints on each axis quantity to give a sense 
of separation between the curves. 
Solid curves 
are for flat space, dotted curves include $\omk$, at the 
mean values from the results tables. 
} 
\label{fig:conjoin} 
\end{figure}

Figure~\ref{fig:conjoin} presents the conjoint history 
diagram for \lcdm\  and the various VM expansion models 
with the parameters given in the Monte Carlo results tables 
for CMB+BAO+SN. 
We see that if we look only at $H_0$ then the VM cases 
are well able to match the high value of $H_0$. Even if 
we also look at $\fsig(z=0)$ (or nearly equivalently $S_8$), 
the  values can be quite close to \lcdm. However, at 
higher redshift the conjoint history deviates sharply, 
and would not be consistent with data agreeing with \lcdm. 
This emphasizes the need for a model to provide a good fit 
to {\it all\/} the data, not just one parameter. We have 
seen this in terms of the poor $\chi^2$ in Sec.~\ref{sec:data} 
and the conjoint history analysis provides another 
view of this.

\section{Discussion and Conclusions} \label{sec:concl} 

Obtaining a value of $H_0\approx74\,$km/s/Mpc from the CMB 
is super easy, barely an inconvenience. Obtaining consistency 
when accounting for data sets from multiple cosmological probes 
-- at a minimum including the recent universe, mid redshifts as 
seen by BAO and SN, and high redshift as evident in the CMB -- 
and ideally both expansion and growth constraints, is not. 
Even then, while 
$\sige$ or $S_8$ can provide a further critical test, these are 
simply the $z=0$ value, and a more incisive test would be consistency 
in a conjoined analysis of both the expansion history and the 
growth history. 

We demonstrate this with the vacuum metamorphosis model, a 
well motivated fundamental physics theory that has properties 
in common with many late time dark energy transition models. 
VM does quite well in fitting the CMB ($\Delta\chi^2\approx -4.9$ to $+2.7$ with respect to \lcdm, with the 
same or one more parameter), and in our baseline combination 
CMB+BAO+SN it achieves $H_0\approx73$--74 without using an $H_0$ 
prior, and gives comparable uncertainty to the \lcdm\ case. 
However, for this three probe combination its goodness of fit 
is poor. That is, looking beyond just $H_0$ the theory fares 
poorly. 

Such fit problems when taking into account the fuller array of 
data is common to many late time phantom dark energy transition attempts to attain  
$H_0\gtrsim70$ (see \cite{DiValentino:2017zyq} for a clear 
illustration of the tension in CMB+BAO+SN when pushing into this 
phantom regime). Adding a vacuum expectation value to VM theory, i.e.\ 
VM VEV, delivers a strong improvement: $\Delta\chi^2=-53$ for 
the one extra parameter, but still insufficient when considering 
the three probe combination. 
We also explore whether including spatial curvature could 
alleviate this, and indeed it can provide strong improvements for 
the CMB+SN case, but not for CMB+BAO+SN. 
Many data combinations prefer a closed universe, at 95\% CL or 
greater, though for the better fitting VM VEV theory and the 
three probe combination, this is reduced to 68\% CL. 

Considering the amplitude of large scale structure, both late 
and early time transition models, by the nature of their mechanism 
for raising $H_0$, also tend to raise $\sige$ or $S_8$ 
above the \lcdm\ value, which is already somewhat high with 
respect to observations. The VM VEV model (or its equivalent 
dark energy expansion behavior) can deliver 
$H_0\approx74$ and a similar $S_8$ to \lcdm\ (though a higher 
$\sige$). However, this is merely a snapshot at $z=0$. We 
emphasize the importance of a full conjoined analysis of the 
expansion and growth histories, by tracing $H(z)$ and $\fsig(z)$ 
at mid redshifts as well. 

In summary, if one has a very narrow focus, e.g.\ just on $H_0$, 
then one can draw a very different conclusion regarding the 
attraction of models than if one properly takes into account the 
array of available data. $H_0$ {\it ex machina\/}, where a model 
swoops in to resolve a seemingly hopeless problem, can be 
intriguing, but given the array of cosmological data we want, and 
need, to go beyond $H_0$. So far, neither early nor late time 
transitions have shown wholly viable solutions to the full 
cosmology.

\acknowledgments
E.D.V.\ acknowledges support from the European Research Council in the form of a Consolidator Grant with number 681431.  E.L.\ is supported in part by the Energetic Cosmos Laboratory and by the 
U.S.\ Department of Energy, Office of Science, Office of High Energy Physics, 
under Award DE-SC-0007867 and contract no.\ DE-AC02-05CH11231. 

\bibliography{biblio}

\begin{thebibliography}{76}%
\makeatletter
\providecommand \@ifxundefined [1]{%
 \@ifx{#1\undefined}
}%
\providecommand \@ifnum [1]{%
 \ifnum #1\expandafter \@firstoftwo
 \else \expandafter \@secondoftwo
 \fi
}%
\providecommand \@ifx [1]{%
 \ifx #1\expandafter \@firstoftwo
 \else \expandafter \@secondoftwo
 \fi
}%
\providecommand \natexlab [1]{#1}%
\providecommand \enquote  [1]{``#1''}%
\providecommand \bibnamefont  [1]{#1}%
\providecommand \bibfnamefont [1]{#1}%
\providecommand \citenamefont [1]{#1}%
\providecommand \href@noop [0]{\@secondoftwo}%
\providecommand \href [0]{\begingroup \@sanitize@url \@href}%
\providecommand \@href[1]{\@@startlink{#1}\@@href}%
\providecommand \@@href[1]{\endgroup#1\@@endlink}%
\providecommand \@sanitize@url [0]{\catcode `\\12\catcode `\$12\catcode
  `\&12\catcode `\#12\catcode `\^12\catcode `\_12\catcode `\%12\relax}%
\providecommand \@@startlink[1]{}%
\providecommand \@@endlink[0]{}%
\providecommand \url  [0]{\begingroup\@sanitize@url \@url }%
\providecommand \@url [1]{\endgroup\@href {#1}{\urlprefix }}%
\providecommand \urlprefix  [0]{URL }%
\providecommand \Eprint [0]{\href }%
\providecommand \doibase [0]{http://dx.doi.org/}%
\providecommand \selectlanguage [0]{\@gobble}%
\providecommand \bibinfo  [0]{\@secondoftwo}%
\providecommand \bibfield  [0]{\@secondoftwo}%
\providecommand \translation [1]{[#1]}%
\providecommand \BibitemOpen [0]{}%
\providecommand \bibitemStop [0]{}%
\providecommand \bibitemNoStop [0]{.\EOS\space}%
\providecommand \EOS [0]{\spacefactor3000\relax}%
\providecommand \BibitemShut  [1]{\csname bibitem#1\endcsname}%
\let\auto@bib@innerbib\@empty
\bibitem [{\citenamefont {Riess}\ \emph {et~al.}(2019)\citenamefont {Riess},
  \citenamefont {Casertano}, \citenamefont {Yuan}, \citenamefont {Macri},\ and\
  \citenamefont {Scolnic}}]{Riess:2019cxk}%
  \BibitemOpen
  \bibfield  {author} {\bibinfo {author} {\bibfnamefont {A.~G.}\ \bibnamefont
  {Riess}}, \bibinfo {author} {\bibfnamefont {S.}~\bibnamefont {Casertano}},
  \bibinfo {author} {\bibfnamefont {W.}~\bibnamefont {Yuan}}, \bibinfo {author}
  {\bibfnamefont {L.~M.}\ \bibnamefont {Macri}}, \ and\ \bibinfo {author}
  {\bibfnamefont {D.}~\bibnamefont {Scolnic}},\ }\href {\doibase
  10.3847/1538-4357/ab1422} {\bibfield  {journal} {\bibinfo  {journal}
  {Astrophys. J.}\ }\textbf {\bibinfo {volume} {876}},\ \bibinfo {pages} {85}
  (\bibinfo {year} {2019})},\ \Eprint {http://arxiv.org/abs/1903.07603}
  {arXiv:1903.07603 [astro-ph.CO]} \BibitemShut {NoStop}%
\bibitem [{\citenamefont {Freedman}\ \emph {et~al.}(2020)\citenamefont
  {Freedman}, \citenamefont {Madore}, \citenamefont {Hoyt}, \citenamefont
  {Jang}, \citenamefont {Beaton}, \citenamefont {Lee}, \citenamefont {Monson},
  \citenamefont {Neeley},\ and\ \citenamefont {Rich}}]{Freedman:2020dne}%
  \BibitemOpen
  \bibfield  {author} {\bibinfo {author} {\bibfnamefont {W.~L.}\ \bibnamefont
  {Freedman}}, \bibinfo {author} {\bibfnamefont {B.~F.}\ \bibnamefont
  {Madore}}, \bibinfo {author} {\bibfnamefont {T.}~\bibnamefont {Hoyt}},
  \bibinfo {author} {\bibfnamefont {I.~S.}\ \bibnamefont {Jang}}, \bibinfo
  {author} {\bibfnamefont {R.}~\bibnamefont {Beaton}}, \bibinfo {author}
  {\bibfnamefont {M.~G.}\ \bibnamefont {Lee}}, \bibinfo {author} {\bibfnamefont
  {A.}~\bibnamefont {Monson}}, \bibinfo {author} {\bibfnamefont
  {J.}~\bibnamefont {Neeley}}, \ and\ \bibinfo {author} {\bibfnamefont
  {J.}~\bibnamefont {Rich}},\ }\href {\doibase 10.3847/1538-4357/ab7339}
  {\bibfield  {journal} {\bibinfo  {journal} {\apj}\ }\textbf {\bibinfo
  {volume} {891}},\ \bibinfo {eid} {57} (\bibinfo {year} {2020})},\ \Eprint
  {http://arxiv.org/abs/2002.01550} {arXiv:2002.01550 [astro-ph.GA]}
  \BibitemShut {NoStop}%
\bibitem [{\citenamefont {Aghanim}\ \emph
  {et~al.}(2018{\natexlab{a}})\citenamefont {Aghanim} \emph
  {et~al.}}]{Aghanim:2018eyx}%
  \BibitemOpen
  \bibfield  {author} {\bibinfo {author} {\bibfnamefont {N.}~\bibnamefont
  {Aghanim}} \emph {et~al.} (\bibinfo {collaboration} {Planck}),\ }\href@noop
  {} {\  (\bibinfo {year} {2018}{\natexlab{a}})},\ \Eprint
  {http://arxiv.org/abs/1807.06209} {arXiv:1807.06209 [astro-ph.CO]}
  \BibitemShut {NoStop}%
\bibitem [{\citenamefont {Bianchini}\ \emph {et~al.}(2020)\citenamefont
  {Bianchini} \emph {et~al.}}]{Bianchini:2019vxp}%
  \BibitemOpen
  \bibfield  {author} {\bibinfo {author} {\bibfnamefont {F.}~\bibnamefont
  {Bianchini}} \emph {et~al.} (\bibinfo {collaboration} {SPT}),\ }\href
  {\doibase 10.3847/1538-4357/ab6082} {\bibfield  {journal} {\bibinfo
  {journal} {Astrophys. J.}\ }\textbf {\bibinfo {volume} {888}},\ \bibinfo
  {pages} {119} (\bibinfo {year} {2020})},\ \Eprint
  {http://arxiv.org/abs/1910.07157} {arXiv:1910.07157 [astro-ph.CO]}
  \BibitemShut {NoStop}%
\bibitem [{\citenamefont {Addison}\ \emph {et~al.}(2018)\citenamefont
  {Addison}, \citenamefont {Watts}, \citenamefont {Bennett}, \citenamefont
  {Halpern}, \citenamefont {Hinshaw},\ and\ \citenamefont
  {Weiland}}]{Addison:2017fdm}%
  \BibitemOpen
  \bibfield  {author} {\bibinfo {author} {\bibfnamefont {G.}~\bibnamefont
  {Addison}}, \bibinfo {author} {\bibfnamefont {D.}~\bibnamefont {Watts}},
  \bibinfo {author} {\bibfnamefont {C.}~\bibnamefont {Bennett}}, \bibinfo
  {author} {\bibfnamefont {M.}~\bibnamefont {Halpern}}, \bibinfo {author}
  {\bibfnamefont {G.}~\bibnamefont {Hinshaw}}, \ and\ \bibinfo {author}
  {\bibfnamefont {J.}~\bibnamefont {Weiland}},\ }\href {\doibase
  10.3847/1538-4357/aaa1ed} {\bibfield  {journal} {\bibinfo  {journal}
  {Astrophys. J.}\ }\textbf {\bibinfo {volume} {853}},\ \bibinfo {pages} {119}
  (\bibinfo {year} {2018})},\ \Eprint {http://arxiv.org/abs/1707.06547}
  {arXiv:1707.06547 [astro-ph.CO]} \BibitemShut {NoStop}%
\bibitem [{\citenamefont {Macaulay}\ \emph {et~al.}(2019)\citenamefont
  {Macaulay} \emph {et~al.}}]{Macaulay:2018fxi}%
  \BibitemOpen
  \bibfield  {author} {\bibinfo {author} {\bibfnamefont {E.}~\bibnamefont
  {Macaulay}} \emph {et~al.} (\bibinfo {collaboration} {DES}),\ }\href
  {\doibase 10.1093/mnras/stz978} {\bibfield  {journal} {\bibinfo  {journal}
  {Mon. Not. Roy. Astron. Soc.}\ }\textbf {\bibinfo {volume} {486}},\ \bibinfo
  {pages} {2184} (\bibinfo {year} {2019})},\ \Eprint
  {http://arxiv.org/abs/1811.02376} {arXiv:1811.02376 [astro-ph.CO]}
  \BibitemShut {NoStop}%
\bibitem [{\citenamefont {Cuceu}\ \emph {et~al.}(2019)\citenamefont {Cuceu},
  \citenamefont {Farr}, \citenamefont {Lemos},\ and\ \citenamefont
  {Font-Ribera}}]{Cuceu:2019for}%
  \BibitemOpen
  \bibfield  {author} {\bibinfo {author} {\bibfnamefont {A.}~\bibnamefont
  {Cuceu}}, \bibinfo {author} {\bibfnamefont {J.}~\bibnamefont {Farr}},
  \bibinfo {author} {\bibfnamefont {P.}~\bibnamefont {Lemos}}, \ and\ \bibinfo
  {author} {\bibfnamefont {A.}~\bibnamefont {Font-Ribera}},\ }\href {\doibase
  10.1088/1475-7516/2019/10/044} {\bibfield  {journal} {\bibinfo  {journal}
  {JCAP}\ }\textbf {\bibinfo {volume} {10}},\ \bibinfo {pages} {044} (\bibinfo
  {year} {2019})},\ \Eprint {http://arxiv.org/abs/1906.11628} {arXiv:1906.11628
  [astro-ph.CO]} \BibitemShut {NoStop}%
\bibitem [{\citenamefont {Wong}\ \emph {et~al.}(2019)\citenamefont {Wong} \emph
  {et~al.}}]{Wong:2019kwg}%
  \BibitemOpen
  \bibfield  {author} {\bibinfo {author} {\bibfnamefont {K.~C.}\ \bibnamefont
  {Wong}} \emph {et~al.},\ }\href {\doibase 10.1093/mnras/stz3094} {\
  (\bibinfo {year} {2019}),\ 10.1093/mnras/stz3094},\ \Eprint
  {http://arxiv.org/abs/1907.04869} {arXiv:1907.04869 [astro-ph.CO]}
  \BibitemShut {NoStop}%
\bibitem [{\citenamefont {Efstathiou}\ and\ \citenamefont
  {Bond}(1999)}]{Efstathiou:1998xx}%
  \BibitemOpen
  \bibfield  {author} {\bibinfo {author} {\bibfnamefont {G.}~\bibnamefont
  {Efstathiou}}\ and\ \bibinfo {author} {\bibfnamefont {J.}~\bibnamefont
  {Bond}},\ }\href {\doibase 10.1046/j.1365-8711.1999.02274.x} {\bibfield
  {journal} {\bibinfo  {journal} {Mon. Not. Roy. Astron. Soc.}\ }\textbf
  {\bibinfo {volume} {304}},\ \bibinfo {pages} {75} (\bibinfo {year} {1999})},\
  \Eprint {http://arxiv.org/abs/astro-ph/9807103} {arXiv:astro-ph/9807103}
  \BibitemShut {NoStop}%
\bibitem [{\citenamefont {Eisenstein}\ and\ \citenamefont
  {White}(2004)}]{Eisenstein:2004an}%
  \BibitemOpen
  \bibfield  {author} {\bibinfo {author} {\bibfnamefont {D.~J.}\ \bibnamefont
  {Eisenstein}}\ and\ \bibinfo {author} {\bibfnamefont {M.~J.}\ \bibnamefont
  {White}},\ }\href {\doibase 10.1103/PhysRevD.70.103523} {\bibfield  {journal}
  {\bibinfo  {journal} {Phys. Rev. D}\ }\textbf {\bibinfo {volume} {70}},\
  \bibinfo {pages} {103523} (\bibinfo {year} {2004})},\ \Eprint
  {http://arxiv.org/abs/astro-ph/0407539} {arXiv:astro-ph/0407539} \BibitemShut
  {NoStop}%
\bibitem [{\citenamefont {Doran}\ \emph {et~al.}(2007)\citenamefont {Doran},
  \citenamefont {Stern},\ and\ \citenamefont {Thommes}}]{Doran:2006xp}%
  \BibitemOpen
  \bibfield  {author} {\bibinfo {author} {\bibfnamefont {M.}~\bibnamefont
  {Doran}}, \bibinfo {author} {\bibfnamefont {S.}~\bibnamefont {Stern}}, \ and\
  \bibinfo {author} {\bibfnamefont {E.}~\bibnamefont {Thommes}},\ }\href
  {\doibase 10.1088/1475-7516/2007/04/015} {\bibfield  {journal} {\bibinfo
  {journal} {JCAP}\ }\textbf {\bibinfo {volume} {04}},\ \bibinfo {pages} {015}
  (\bibinfo {year} {2007})},\ \Eprint {http://arxiv.org/abs/astro-ph/0609075}
  {arXiv:astro-ph/0609075} \BibitemShut {NoStop}%
\bibitem [{\citenamefont {Linder}\ and\ \citenamefont
  {Robbers}(2008)}]{Linder:2008nq}%
  \BibitemOpen
  \bibfield  {author} {\bibinfo {author} {\bibfnamefont {E.~V.}\ \bibnamefont
  {Linder}}\ and\ \bibinfo {author} {\bibfnamefont {G.}~\bibnamefont
  {Robbers}},\ }\href {\doibase 10.1088/1475-7516/2008/06/004} {\bibfield
  {journal} {\bibinfo  {journal} {JCAP}\ }\textbf {\bibinfo {volume} {06}},\
  \bibinfo {pages} {004} (\bibinfo {year} {2008})},\ \Eprint
  {http://arxiv.org/abs/0803.2877} {arXiv:0803.2877 [astro-ph]} \BibitemShut
  {NoStop}%
\bibitem [{\citenamefont {Hojjati}\ \emph {et~al.}(2013)\citenamefont
  {Hojjati}, \citenamefont {Linder},\ and\ \citenamefont
  {Samsing}}]{Hojjati:2013oya}%
  \BibitemOpen
  \bibfield  {author} {\bibinfo {author} {\bibfnamefont {A.}~\bibnamefont
  {Hojjati}}, \bibinfo {author} {\bibfnamefont {E.~V.}\ \bibnamefont {Linder}},
  \ and\ \bibinfo {author} {\bibfnamefont {J.}~\bibnamefont {Samsing}},\ }\href
  {\doibase 10.1103/PhysRevLett.111.041301} {\bibfield  {journal} {\bibinfo
  {journal} {Phys. Rev. Lett.}\ }\textbf {\bibinfo {volume} {111}},\ \bibinfo
  {pages} {041301} (\bibinfo {year} {2013})},\ \Eprint
  {http://arxiv.org/abs/1304.3724} {arXiv:1304.3724 [astro-ph.CO]} \BibitemShut
  {NoStop}%
\bibitem [{\citenamefont {Di~Valentino}\ \emph {et~al.}(2018)\citenamefont
  {Di~Valentino}, \citenamefont {Linder},\ and\ \citenamefont
  {Melchiorri}}]{DiValentino:2017rcr}%
  \BibitemOpen
  \bibfield  {author} {\bibinfo {author} {\bibfnamefont {E.}~\bibnamefont
  {Di~Valentino}}, \bibinfo {author} {\bibfnamefont {E.~V.}\ \bibnamefont
  {Linder}}, \ and\ \bibinfo {author} {\bibfnamefont {A.}~\bibnamefont
  {Melchiorri}},\ }\href {\doibase 10.1103/PhysRevD.97.043528} {\bibfield
  {journal} {\bibinfo  {journal} {Phys. Rev. D}\ }\textbf {\bibinfo {volume}
  {97}},\ \bibinfo {pages} {043528} (\bibinfo {year} {2018})},\ \Eprint
  {http://arxiv.org/abs/1710.02153} {arXiv:1710.02153 [astro-ph.CO]}
  \BibitemShut {NoStop}%
\bibitem [{\citenamefont {Li}\ and\ \citenamefont
  {Shafieloo}(2019)}]{Li:2019yem}%
  \BibitemOpen
  \bibfield  {author} {\bibinfo {author} {\bibfnamefont {X.}~\bibnamefont
  {Li}}\ and\ \bibinfo {author} {\bibfnamefont {A.}~\bibnamefont {Shafieloo}},\
  }\href {\doibase 10.3847/2041-8213/ab3e09} {\bibfield  {journal} {\bibinfo
  {journal} {Astrophys. J. Lett.}\ }\textbf {\bibinfo {volume} {883}},\
  \bibinfo {pages} {L3} (\bibinfo {year} {2019})},\ \Eprint
  {http://arxiv.org/abs/1906.08275} {arXiv:1906.08275 [astro-ph.CO]}
  \BibitemShut {NoStop}%
\bibitem [{\citenamefont {Pan}\ \emph {et~al.}(2019)\citenamefont {Pan},
  \citenamefont {Yang}, \citenamefont {Di~Valentino}, \citenamefont
  {Shafieloo},\ and\ \citenamefont {Chakraborty}}]{Pan:2019hac}%
  \BibitemOpen
  \bibfield  {author} {\bibinfo {author} {\bibfnamefont {S.}~\bibnamefont
  {Pan}}, \bibinfo {author} {\bibfnamefont {W.}~\bibnamefont {Yang}}, \bibinfo
  {author} {\bibfnamefont {E.}~\bibnamefont {Di~Valentino}}, \bibinfo {author}
  {\bibfnamefont {A.}~\bibnamefont {Shafieloo}}, \ and\ \bibinfo {author}
  {\bibfnamefont {S.}~\bibnamefont {Chakraborty}},\ }\href@noop {} {\
  (\bibinfo {year} {2019})},\ \Eprint {http://arxiv.org/abs/1907.12551}
  {arXiv:1907.12551 [astro-ph.CO]} \BibitemShut {NoStop}%
\bibitem [{\citenamefont {Li}\ and\ \citenamefont
  {Shafieloo}(2020)}]{Li:2020ybr}%
  \BibitemOpen
  \bibfield  {author} {\bibinfo {author} {\bibfnamefont {X.}~\bibnamefont
  {Li}}\ and\ \bibinfo {author} {\bibfnamefont {A.}~\bibnamefont {Shafieloo}},\
  }\href@noop {} {\  (\bibinfo {year} {2020})},\ \Eprint
  {http://arxiv.org/abs/2001.05103} {arXiv:2001.05103 [astro-ph.CO]}
  \BibitemShut {NoStop}%
\bibitem [{\citenamefont {Khosravi}\ \emph {et~al.}(2019)\citenamefont
  {Khosravi}, \citenamefont {Baghram}, \citenamefont {Afshordi},\ and\
  \citenamefont {Altamirano}}]{Khosravi:2017hfi}%
  \BibitemOpen
  \bibfield  {author} {\bibinfo {author} {\bibfnamefont {N.}~\bibnamefont
  {Khosravi}}, \bibinfo {author} {\bibfnamefont {S.}~\bibnamefont {Baghram}},
  \bibinfo {author} {\bibfnamefont {N.}~\bibnamefont {Afshordi}}, \ and\
  \bibinfo {author} {\bibfnamefont {N.}~\bibnamefont {Altamirano}},\ }\href
  {\doibase 10.1103/PhysRevD.99.103526} {\bibfield  {journal} {\bibinfo
  {journal} {Phys. Rev. D}\ }\textbf {\bibinfo {volume} {99}},\ \bibinfo
  {pages} {103526} (\bibinfo {year} {2019})},\ \Eprint
  {http://arxiv.org/abs/1710.09366} {arXiv:1710.09366 [astro-ph.CO]}
  \BibitemShut {NoStop}%
\bibitem [{\citenamefont {Di~Valentino}\ \emph
  {et~al.}(2020{\natexlab{a}})\citenamefont {Di~Valentino}, \citenamefont
  {Mukherjee},\ and\ \citenamefont {Sen}}]{DiValentino:2020naf}%
  \BibitemOpen
  \bibfield  {author} {\bibinfo {author} {\bibfnamefont {E.}~\bibnamefont
  {Di~Valentino}}, \bibinfo {author} {\bibfnamefont {A.}~\bibnamefont
  {Mukherjee}}, \ and\ \bibinfo {author} {\bibfnamefont {A.~A.}\ \bibnamefont
  {Sen}},\ }\href@noop {} {\  (\bibinfo {year} {2020}{\natexlab{a}})},\ \Eprint
  {http://arxiv.org/abs/2005.12587} {arXiv:2005.12587 [astro-ph.CO]}
  \BibitemShut {NoStop}%
\bibitem [{\citenamefont {Poulin}\ \emph {et~al.}(2019)\citenamefont {Poulin},
  \citenamefont {Smith}, \citenamefont {Karwal},\ and\ \citenamefont
  {Kamionkowski}}]{Poulin:2018cxd}%
  \BibitemOpen
  \bibfield  {author} {\bibinfo {author} {\bibfnamefont {V.}~\bibnamefont
  {Poulin}}, \bibinfo {author} {\bibfnamefont {T.~L.}\ \bibnamefont {Smith}},
  \bibinfo {author} {\bibfnamefont {T.}~\bibnamefont {Karwal}}, \ and\ \bibinfo
  {author} {\bibfnamefont {M.}~\bibnamefont {Kamionkowski}},\ }\href {\doibase
  10.1103/PhysRevLett.122.221301} {\bibfield  {journal} {\bibinfo  {journal}
  {Phys. Rev. Lett.}\ }\textbf {\bibinfo {volume} {122}},\ \bibinfo {pages}
  {221301} (\bibinfo {year} {2019})},\ \Eprint
  {http://arxiv.org/abs/1811.04083} {arXiv:1811.04083 [astro-ph.CO]}
  \BibitemShut {NoStop}%
\bibitem [{\citenamefont {Smith}\ \emph {et~al.}(2020)\citenamefont {Smith},
  \citenamefont {Poulin},\ and\ \citenamefont {Amin}}]{Smith:2019ihp}%
  \BibitemOpen
  \bibfield  {author} {\bibinfo {author} {\bibfnamefont {T.~L.}\ \bibnamefont
  {Smith}}, \bibinfo {author} {\bibfnamefont {V.}~\bibnamefont {Poulin}}, \
  and\ \bibinfo {author} {\bibfnamefont {M.~A.}\ \bibnamefont {Amin}},\ }\href
  {\doibase 10.1103/PhysRevD.101.063523} {\bibfield  {journal} {\bibinfo
  {journal} {Phys. Rev. D}\ }\textbf {\bibinfo {volume} {101}},\ \bibinfo
  {pages} {063523} (\bibinfo {year} {2020})},\ \Eprint
  {http://arxiv.org/abs/1908.06995} {arXiv:1908.06995 [astro-ph.CO]}
  \BibitemShut {NoStop}%
\bibitem [{\citenamefont {Agrawal}\ \emph {et~al.}(2019)\citenamefont
  {Agrawal}, \citenamefont {Cyr-Racine}, \citenamefont {Pinner},\ and\
  \citenamefont {Randall}}]{Agrawal:2019lmo}%
  \BibitemOpen
  \bibfield  {author} {\bibinfo {author} {\bibfnamefont {P.}~\bibnamefont
  {Agrawal}}, \bibinfo {author} {\bibfnamefont {F.-Y.}\ \bibnamefont
  {Cyr-Racine}}, \bibinfo {author} {\bibfnamefont {D.}~\bibnamefont {Pinner}},
  \ and\ \bibinfo {author} {\bibfnamefont {L.}~\bibnamefont {Randall}},\
  }\href@noop {} {\  (\bibinfo {year} {2019})},\ \Eprint
  {http://arxiv.org/abs/1904.01016} {arXiv:1904.01016 [astro-ph.CO]}
  \BibitemShut {NoStop}%
\bibitem [{\citenamefont {Lin}\ \emph {et~al.}(2019)\citenamefont {Lin},
  \citenamefont {Benevento}, \citenamefont {Hu},\ and\ \citenamefont
  {Raveri}}]{Lin:2019qug}%
  \BibitemOpen
  \bibfield  {author} {\bibinfo {author} {\bibfnamefont {M.-X.}\ \bibnamefont
  {Lin}}, \bibinfo {author} {\bibfnamefont {G.}~\bibnamefont {Benevento}},
  \bibinfo {author} {\bibfnamefont {W.}~\bibnamefont {Hu}}, \ and\ \bibinfo
  {author} {\bibfnamefont {M.}~\bibnamefont {Raveri}},\ }\href {\doibase
  10.1103/PhysRevD.100.063542} {\bibfield  {journal} {\bibinfo  {journal}
  {Phys. Rev. D}\ }\textbf {\bibinfo {volume} {100}},\ \bibinfo {pages}
  {063542} (\bibinfo {year} {2019})},\ \Eprint
  {http://arxiv.org/abs/1905.12618} {arXiv:1905.12618 [astro-ph.CO]}
  \BibitemShut {NoStop}%
\bibitem [{\citenamefont {Niedermann}\ and\ \citenamefont
  {Sloth}(2019)}]{Niedermann:2019olb}%
  \BibitemOpen
  \bibfield  {author} {\bibinfo {author} {\bibfnamefont {F.}~\bibnamefont
  {Niedermann}}\ and\ \bibinfo {author} {\bibfnamefont {M.~S.}\ \bibnamefont
  {Sloth}},\ }\href@noop {} {\  (\bibinfo {year} {2019})},\ \Eprint
  {http://arxiv.org/abs/1910.10739} {arXiv:1910.10739 [astro-ph.CO]}
  \BibitemShut {NoStop}%
\bibitem [{\citenamefont {Hill}\ \emph {et~al.}(2020)\citenamefont {Hill},
  \citenamefont {McDonough}, \citenamefont {Toomey},\ and\ \citenamefont
  {Alexander}}]{Hill:2020osr}%
  \BibitemOpen
  \bibfield  {author} {\bibinfo {author} {\bibfnamefont {J.~C.}\ \bibnamefont
  {Hill}}, \bibinfo {author} {\bibfnamefont {E.}~\bibnamefont {McDonough}},
  \bibinfo {author} {\bibfnamefont {M.~W.}\ \bibnamefont {Toomey}}, \ and\
  \bibinfo {author} {\bibfnamefont {S.}~\bibnamefont {Alexander}},\ }\href@noop
  {} {\  (\bibinfo {year} {2020})},\ \Eprint {http://arxiv.org/abs/2003.07355}
  {arXiv:2003.07355 [astro-ph.CO]} \BibitemShut {NoStop}%
\bibitem [{\citenamefont {Benevento}\ \emph {et~al.}(2020)\citenamefont
  {Benevento}, \citenamefont {Hu},\ and\ \citenamefont
  {Raveri}}]{Benevento:2020fev}%
  \BibitemOpen
  \bibfield  {author} {\bibinfo {author} {\bibfnamefont {G.}~\bibnamefont
  {Benevento}}, \bibinfo {author} {\bibfnamefont {W.}~\bibnamefont {Hu}}, \
  and\ \bibinfo {author} {\bibfnamefont {M.}~\bibnamefont {Raveri}},\ }\href
  {\doibase 10.1103/PhysRevD.101.103517} {\bibfield  {journal} {\bibinfo
  {journal} {Phys. Rev. D}\ }\textbf {\bibinfo {volume} {101}},\ \bibinfo
  {pages} {103517} (\bibinfo {year} {2020})},\ \Eprint
  {http://arxiv.org/abs/2002.11707} {arXiv:2002.11707 [astro-ph.CO]}
  \BibitemShut {NoStop}%
\bibitem [{\citenamefont {Knox}\ and\ \citenamefont
  {Millea}(2020)}]{Knox:2019rjx}%
  \BibitemOpen
  \bibfield  {author} {\bibinfo {author} {\bibfnamefont {L.}~\bibnamefont
  {Knox}}\ and\ \bibinfo {author} {\bibfnamefont {M.}~\bibnamefont {Millea}},\
  }\href {\doibase 10.1103/PhysRevD.101.043533} {\bibfield  {journal} {\bibinfo
   {journal} {Phys. Rev. D}\ }\textbf {\bibinfo {volume} {101}},\ \bibinfo
  {pages} {043533} (\bibinfo {year} {2020})},\ \Eprint
  {http://arxiv.org/abs/1908.03663} {arXiv:1908.03663 [astro-ph.CO]}
  \BibitemShut {NoStop}%
\bibitem [{\citenamefont {Di~Valentino}\ \emph
  {et~al.}(2019{\natexlab{a}})\citenamefont {Di~Valentino}, \citenamefont
  {Melchiorri}, \citenamefont {Mena},\ and\ \citenamefont
  {Vagnozzi}}]{DiValentino:2019ffd}%
  \BibitemOpen
  \bibfield  {author} {\bibinfo {author} {\bibfnamefont {E.}~\bibnamefont
  {Di~Valentino}}, \bibinfo {author} {\bibfnamefont {A.}~\bibnamefont
  {Melchiorri}}, \bibinfo {author} {\bibfnamefont {O.}~\bibnamefont {Mena}}, \
  and\ \bibinfo {author} {\bibfnamefont {S.}~\bibnamefont {Vagnozzi}},\
  }\href@noop {} {\  (\bibinfo {year} {2019}{\natexlab{a}})},\ \Eprint
  {http://arxiv.org/abs/1908.04281} {arXiv:1908.04281 [astro-ph.CO]}
  \BibitemShut {NoStop}%
\bibitem [{\citenamefont {Arendse}\ \emph {et~al.}(2019)\citenamefont {Arendse}
  \emph {et~al.}}]{Arendse:2019hev}%
  \BibitemOpen
  \bibfield  {author} {\bibinfo {author} {\bibfnamefont {N.}~\bibnamefont
  {Arendse}} \emph {et~al.},\ }\href@noop {} {\  (\bibinfo {year} {2019})},\
  \Eprint {http://arxiv.org/abs/1909.07986} {arXiv:1909.07986 [astro-ph.CO]}
  \BibitemShut {NoStop}%
\bibitem [{\citenamefont {Garcia-Quintero}\ \emph {et~al.}(2019)\citenamefont
  {Garcia-Quintero}, \citenamefont {Ishak}, \citenamefont {Fox},\ and\
  \citenamefont {Lin}}]{Garcia-Quintero:2019cgt}%
  \BibitemOpen
  \bibfield  {author} {\bibinfo {author} {\bibfnamefont {C.}~\bibnamefont
  {Garcia-Quintero}}, \bibinfo {author} {\bibfnamefont {M.}~\bibnamefont
  {Ishak}}, \bibinfo {author} {\bibfnamefont {L.}~\bibnamefont {Fox}}, \ and\
  \bibinfo {author} {\bibfnamefont {W.}~\bibnamefont {Lin}},\ }\href {\doibase
  10.1103/PhysRevD.100.123538} {\bibfield  {journal} {\bibinfo  {journal}
  {Phys. Rev. D}\ }\textbf {\bibinfo {volume} {100}},\ \bibinfo {pages}
  {123538} (\bibinfo {year} {2019})},\ \Eprint
  {http://arxiv.org/abs/1910.01608} {arXiv:1910.01608 [astro-ph.CO]}
  \BibitemShut {NoStop}%
\bibitem [{\citenamefont {Hart}\ and\ \citenamefont
  {Chluba}(2019)}]{Hart:2019dxi}%
  \BibitemOpen
  \bibfield  {author} {\bibinfo {author} {\bibfnamefont {L.}~\bibnamefont
  {Hart}}\ and\ \bibinfo {author} {\bibfnamefont {J.}~\bibnamefont {Chluba}},\
  }\href@noop {} {\  (\bibinfo {year} {2019})},\ \Eprint
  {http://arxiv.org/abs/1912.03986} {arXiv:1912.03986 [astro-ph.CO]}
  \BibitemShut {NoStop}%
\bibitem [{\citenamefont {Di~Valentino}\ \emph
  {et~al.}(2019{\natexlab{b}})\citenamefont {Di~Valentino}, \citenamefont
  {Melchiorri}, \citenamefont {Mena},\ and\ \citenamefont
  {Vagnozzi}}]{DiValentino:2019jae}%
  \BibitemOpen
  \bibfield  {author} {\bibinfo {author} {\bibfnamefont {E.}~\bibnamefont
  {Di~Valentino}}, \bibinfo {author} {\bibfnamefont {A.}~\bibnamefont
  {Melchiorri}}, \bibinfo {author} {\bibfnamefont {O.}~\bibnamefont {Mena}}, \
  and\ \bibinfo {author} {\bibfnamefont {S.}~\bibnamefont {Vagnozzi}},\
  }\href@noop {} {\  (\bibinfo {year} {2019}{\natexlab{b}})},\ \Eprint
  {http://arxiv.org/abs/1910.09853} {arXiv:1910.09853 [astro-ph.CO]}
  \BibitemShut {NoStop}%
\bibitem [{\citenamefont {Liu}\ \emph {et~al.}(2020)\citenamefont {Liu},
  \citenamefont {Huang}, \citenamefont {Luo}, \citenamefont {Miao},
  \citenamefont {Singh},\ and\ \citenamefont {Huang}}]{Liu:2019awo}%
  \BibitemOpen
  \bibfield  {author} {\bibinfo {author} {\bibfnamefont {M.}~\bibnamefont
  {Liu}}, \bibinfo {author} {\bibfnamefont {Z.}~\bibnamefont {Huang}}, \bibinfo
  {author} {\bibfnamefont {X.}~\bibnamefont {Luo}}, \bibinfo {author}
  {\bibfnamefont {H.}~\bibnamefont {Miao}}, \bibinfo {author} {\bibfnamefont
  {N.~K.}\ \bibnamefont {Singh}}, \ and\ \bibinfo {author} {\bibfnamefont
  {L.}~\bibnamefont {Huang}},\ }\href {\doibase 10.1007/s11433-019-1509-5}
  {\bibfield  {journal} {\bibinfo  {journal} {Sci. China Phys. Mech. Astron.}\
  }\textbf {\bibinfo {volume} {63}},\ \bibinfo {pages} {290405} (\bibinfo
  {year} {2020})},\ \Eprint {http://arxiv.org/abs/1912.00190} {arXiv:1912.00190
  [astro-ph.CO]} \BibitemShut {NoStop}%
\bibitem [{\citenamefont {Ivanov}\ \emph {et~al.}(2020)\citenamefont {Ivanov},
  \citenamefont {Simonovi\'c},\ and\ \citenamefont
  {Zaldarriaga}}]{Ivanov:2019hqk}%
  \BibitemOpen
  \bibfield  {author} {\bibinfo {author} {\bibfnamefont {M.~M.}\ \bibnamefont
  {Ivanov}}, \bibinfo {author} {\bibfnamefont {M.}~\bibnamefont {Simonovi\'c}},
  \ and\ \bibinfo {author} {\bibfnamefont {M.}~\bibnamefont {Zaldarriaga}},\
  }\href {\doibase 10.1103/PhysRevD.101.083504} {\bibfield  {journal} {\bibinfo
   {journal} {Phys. Rev. D}\ }\textbf {\bibinfo {volume} {101}},\ \bibinfo
  {pages} {083504} (\bibinfo {year} {2020})},\ \Eprint
  {http://arxiv.org/abs/1912.08208} {arXiv:1912.08208 [astro-ph.CO]}
  \BibitemShut {NoStop}%
\bibitem [{\citenamefont {Alcaniz}\ \emph {et~al.}(2019)\citenamefont
  {Alcaniz}, \citenamefont {Bernal}, \citenamefont {Masiero},\ and\
  \citenamefont {Queiroz}}]{Alcaniz:2019kah}%
  \BibitemOpen
  \bibfield  {author} {\bibinfo {author} {\bibfnamefont {J.}~\bibnamefont
  {Alcaniz}}, \bibinfo {author} {\bibfnamefont {N.}~\bibnamefont {Bernal}},
  \bibinfo {author} {\bibfnamefont {A.}~\bibnamefont {Masiero}}, \ and\
  \bibinfo {author} {\bibfnamefont {F.~S.}\ \bibnamefont {Queiroz}},\
  }\href@noop {} {\  (\bibinfo {year} {2019})},\ \Eprint
  {http://arxiv.org/abs/1912.05563} {arXiv:1912.05563 [astro-ph.CO]}
  \BibitemShut {NoStop}%
\bibitem [{\citenamefont {Frusciante}\ \emph {et~al.}(2019)\citenamefont
  {Frusciante}, \citenamefont {Peirone}, \citenamefont {Atayde},\ and\
  \citenamefont {De~Felice}}]{Frusciante:2019puu}%
  \BibitemOpen
  \bibfield  {author} {\bibinfo {author} {\bibfnamefont {N.}~\bibnamefont
  {Frusciante}}, \bibinfo {author} {\bibfnamefont {S.}~\bibnamefont {Peirone}},
  \bibinfo {author} {\bibfnamefont {L.}~\bibnamefont {Atayde}}, \ and\ \bibinfo
  {author} {\bibfnamefont {A.}~\bibnamefont {De~Felice}},\ }\href@noop {} {\
  (\bibinfo {year} {2019})},\ \Eprint {http://arxiv.org/abs/1912.07586}
  {arXiv:1912.07586 [astro-ph.CO]} \BibitemShut {NoStop}%
\bibitem [{\citenamefont {Yang}\ \emph {et~al.}(2020)\citenamefont {Yang},
  \citenamefont {Di~Valentino}, \citenamefont {Pan}, \citenamefont
  {Basilakos},\ and\ \citenamefont {Paliathanasis}}]{Yang:2020zuk}%
  \BibitemOpen
  \bibfield  {author} {\bibinfo {author} {\bibfnamefont {W.}~\bibnamefont
  {Yang}}, \bibinfo {author} {\bibfnamefont {E.}~\bibnamefont {Di~Valentino}},
  \bibinfo {author} {\bibfnamefont {S.}~\bibnamefont {Pan}}, \bibinfo {author}
  {\bibfnamefont {S.}~\bibnamefont {Basilakos}}, \ and\ \bibinfo {author}
  {\bibfnamefont {A.}~\bibnamefont {Paliathanasis}},\ }\href@noop {} {\
  (\bibinfo {year} {2020})},\ \Eprint {http://arxiv.org/abs/2001.04307}
  {arXiv:2001.04307 [astro-ph.CO]} \BibitemShut {NoStop}%
\bibitem [{\citenamefont {Jedamzik}\ and\ \citenamefont
  {Pogosian}(2020)}]{Jedamzik:2020krr}%
  \BibitemOpen
  \bibfield  {author} {\bibinfo {author} {\bibfnamefont {K.}~\bibnamefont
  {Jedamzik}}\ and\ \bibinfo {author} {\bibfnamefont {L.}~\bibnamefont
  {Pogosian}},\ }\href@noop {} {\  (\bibinfo {year} {2020})},\ \Eprint
  {http://arxiv.org/abs/2004.09487} {arXiv:2004.09487 [astro-ph.CO]}
  \BibitemShut {NoStop}%
\bibitem [{\citenamefont {Pan}\ \emph {et~al.}(2020)\citenamefont {Pan},
  \citenamefont {Sharov},\ and\ \citenamefont {Yang}}]{Pan:2020zza}%
  \BibitemOpen
  \bibfield  {author} {\bibinfo {author} {\bibfnamefont {S.}~\bibnamefont
  {Pan}}, \bibinfo {author} {\bibfnamefont {G.~S.}\ \bibnamefont {Sharov}}, \
  and\ \bibinfo {author} {\bibfnamefont {W.}~\bibnamefont {Yang}},\ }\href@noop
  {} {\  (\bibinfo {year} {2020})},\ \Eprint {http://arxiv.org/abs/2001.03120}
  {arXiv:2001.03120 [astro-ph.CO]} \BibitemShut {NoStop}%
\bibitem [{\citenamefont {Wu}\ \emph {et~al.}(2020)\citenamefont {Wu},
  \citenamefont {Motloch}, \citenamefont {Hu},\ and\ \citenamefont
  {Raveri}}]{Wu:2020nxz}%
  \BibitemOpen
  \bibfield  {author} {\bibinfo {author} {\bibfnamefont {W.~K.}\ \bibnamefont
  {Wu}}, \bibinfo {author} {\bibfnamefont {P.}~\bibnamefont {Motloch}},
  \bibinfo {author} {\bibfnamefont {W.}~\bibnamefont {Hu}}, \ and\ \bibinfo
  {author} {\bibfnamefont {M.}~\bibnamefont {Raveri}},\ }\href@noop {} {\
  (\bibinfo {year} {2020})},\ \Eprint {http://arxiv.org/abs/2004.10207}
  {arXiv:2004.10207 [astro-ph.CO]} \BibitemShut {NoStop}%
\bibitem [{\citenamefont {Ye}\ and\ \citenamefont {Piao}(2020)}]{Ye:2020btb}%
  \BibitemOpen
  \bibfield  {author} {\bibinfo {author} {\bibfnamefont {G.}~\bibnamefont
  {Ye}}\ and\ \bibinfo {author} {\bibfnamefont {Y.-S.}\ \bibnamefont {Piao}},\
  }\href@noop {} {\  (\bibinfo {year} {2020})},\ \Eprint
  {http://arxiv.org/abs/2001.02451} {arXiv:2001.02451 [astro-ph.CO]}
  \BibitemShut {NoStop}%
\bibitem [{\citenamefont {Braglia}\ \emph {et~al.}(2020)\citenamefont
  {Braglia}, \citenamefont {Ballardini}, \citenamefont {Emond}, \citenamefont
  {Finelli}, \citenamefont {Gumrukcuoglu}, \citenamefont {Koyama},\ and\
  \citenamefont {Paoletti}}]{Braglia:2020iik}%
  \BibitemOpen
  \bibfield  {author} {\bibinfo {author} {\bibfnamefont {M.}~\bibnamefont
  {Braglia}}, \bibinfo {author} {\bibfnamefont {M.}~\bibnamefont {Ballardini}},
  \bibinfo {author} {\bibfnamefont {W.~T.}\ \bibnamefont {Emond}}, \bibinfo
  {author} {\bibfnamefont {F.}~\bibnamefont {Finelli}}, \bibinfo {author}
  {\bibfnamefont {A.~E.}\ \bibnamefont {Gumrukcuoglu}}, \bibinfo {author}
  {\bibfnamefont {K.}~\bibnamefont {Koyama}}, \ and\ \bibinfo {author}
  {\bibfnamefont {D.}~\bibnamefont {Paoletti}},\ }\href@noop {} {\  (\bibinfo
  {year} {2020})},\ \Eprint {http://arxiv.org/abs/2004.11161} {arXiv:2004.11161
  [astro-ph.CO]} \BibitemShut {NoStop}%
\bibitem [{\citenamefont {Blinov}\ and\ \citenamefont
  {Marques-Tavares}(2020)}]{Blinov:2020hmc}%
  \BibitemOpen
  \bibfield  {author} {\bibinfo {author} {\bibfnamefont {N.}~\bibnamefont
  {Blinov}}\ and\ \bibinfo {author} {\bibfnamefont {G.}~\bibnamefont
  {Marques-Tavares}},\ }\href@noop {} {\  (\bibinfo {year} {2020})},\ \Eprint
  {http://arxiv.org/abs/2003.08387} {arXiv:2003.08387 [astro-ph.CO]}
  \BibitemShut {NoStop}%
\bibitem [{\citenamefont {Wang}\ and\ \citenamefont
  {Mota}(2020)}]{Wang:2020zfv}%
  \BibitemOpen
  \bibfield  {author} {\bibinfo {author} {\bibfnamefont {D.}~\bibnamefont
  {Wang}}\ and\ \bibinfo {author} {\bibfnamefont {D.}~\bibnamefont {Mota}},\
  }\href@noop {} {\  (\bibinfo {year} {2020})},\ \Eprint
  {http://arxiv.org/abs/2003.10095} {arXiv:2003.10095 [astro-ph.CO]}
  \BibitemShut {NoStop}%
\bibitem [{\citenamefont {Chudaykin}\ \emph {et~al.}(2020)\citenamefont
  {Chudaykin}, \citenamefont {Gorbunov},\ and\ \citenamefont
  {Nedelko}}]{Chudaykin:2020acu}%
  \BibitemOpen
  \bibfield  {author} {\bibinfo {author} {\bibfnamefont {A.}~\bibnamefont
  {Chudaykin}}, \bibinfo {author} {\bibfnamefont {D.}~\bibnamefont {Gorbunov}},
  \ and\ \bibinfo {author} {\bibfnamefont {N.}~\bibnamefont {Nedelko}},\
  }\href@noop {} {\  (\bibinfo {year} {2020})},\ \Eprint
  {http://arxiv.org/abs/2004.13046} {arXiv:2004.13046 [astro-ph.CO]}
  \BibitemShut {NoStop}%
\bibitem [{\citenamefont {Alestas}\ \emph {et~al.}(2020)\citenamefont
  {Alestas}, \citenamefont {Kazantzidis},\ and\ \citenamefont
  {Perivolaropoulos}}]{Alestas:2020mvb}%
  \BibitemOpen
  \bibfield  {author} {\bibinfo {author} {\bibfnamefont {G.}~\bibnamefont
  {Alestas}}, \bibinfo {author} {\bibfnamefont {L.}~\bibnamefont
  {Kazantzidis}}, \ and\ \bibinfo {author} {\bibfnamefont {L.}~\bibnamefont
  {Perivolaropoulos}},\ }\href {\doibase 10.1103/PhysRevD.101.123516}
  {\bibfield  {journal} {\bibinfo  {journal} {Phys. Rev. D}\ }\textbf {\bibinfo
  {volume} {101}},\ \bibinfo {pages} {123516} (\bibinfo {year} {2020})},\
  \Eprint {http://arxiv.org/abs/2004.08363} {arXiv:2004.08363 [astro-ph.CO]}
  \BibitemShut {NoStop}%
\bibitem [{\citenamefont {Clark}\ \emph {et~al.}(2020)\citenamefont {Clark},
  \citenamefont {Vattis},\ and\ \citenamefont {Koushiappas}}]{Clark:2020miy}%
  \BibitemOpen
  \bibfield  {author} {\bibinfo {author} {\bibfnamefont {S.~J.}\ \bibnamefont
  {Clark}}, \bibinfo {author} {\bibfnamefont {K.}~\bibnamefont {Vattis}}, \
  and\ \bibinfo {author} {\bibfnamefont {S.~M.}\ \bibnamefont {Koushiappas}},\
  }\href@noop {} {\  (\bibinfo {year} {2020})},\ \Eprint
  {http://arxiv.org/abs/2006.03678} {arXiv:2006.03678 [astro-ph.CO]}
  \BibitemShut {NoStop}%
\bibitem [{\citenamefont {Ballardini}\ \emph {et~al.}(2020)\citenamefont
  {Ballardini}, \citenamefont {Braglia}, \citenamefont {Finelli}, \citenamefont
  {Paoletti}, \citenamefont {Starobinsky},\ and\ \citenamefont
  {Umiltà}}]{Ballardini:2020iws}%
  \BibitemOpen
  \bibfield  {author} {\bibinfo {author} {\bibfnamefont {M.}~\bibnamefont
  {Ballardini}}, \bibinfo {author} {\bibfnamefont {M.}~\bibnamefont {Braglia}},
  \bibinfo {author} {\bibfnamefont {F.}~\bibnamefont {Finelli}}, \bibinfo
  {author} {\bibfnamefont {D.}~\bibnamefont {Paoletti}}, \bibinfo {author}
  {\bibfnamefont {A.~A.}\ \bibnamefont {Starobinsky}}, \ and\ \bibinfo {author}
  {\bibfnamefont {C.}~\bibnamefont {Umiltà}},\ }\href@noop {} {\  (\bibinfo
  {year} {2020})},\ \Eprint {http://arxiv.org/abs/2004.14349} {arXiv:2004.14349
  [astro-ph.CO]} \BibitemShut {NoStop}%
\bibitem [{\citenamefont {Keeley}\ \emph {et~al.}(2020)\citenamefont {Keeley},
  \citenamefont {Shafieloo}, \citenamefont {Hazra},\ and\ \citenamefont
  {Souradeep}}]{Keeley:2020rmo}%
  \BibitemOpen
  \bibfield  {author} {\bibinfo {author} {\bibfnamefont {R.~E.}\ \bibnamefont
  {Keeley}}, \bibinfo {author} {\bibfnamefont {A.}~\bibnamefont {Shafieloo}},
  \bibinfo {author} {\bibfnamefont {D.~K.}\ \bibnamefont {Hazra}}, \ and\
  \bibinfo {author} {\bibfnamefont {T.}~\bibnamefont {Souradeep}},\ }\href@noop
  {} {\  (\bibinfo {year} {2020})},\ \Eprint {http://arxiv.org/abs/2006.12710}
  {arXiv:2006.12710 [astro-ph.CO]} \BibitemShut {NoStop}%
\bibitem [{\citenamefont {Niedermann}\ and\ \citenamefont
  {Sloth}(2020)}]{Niedermann:2020dwg}%
  \BibitemOpen
  \bibfield  {author} {\bibinfo {author} {\bibfnamefont {F.}~\bibnamefont
  {Niedermann}}\ and\ \bibinfo {author} {\bibfnamefont {M.~S.}\ \bibnamefont
  {Sloth}},\ }\href@noop {} {\  (\bibinfo {year} {2020})},\ \Eprint
  {http://arxiv.org/abs/2006.06686} {arXiv:2006.06686 [astro-ph.CO]}
  \BibitemShut {NoStop}%
\bibitem [{\citenamefont {Archidiacono}\ \emph {et~al.}(2020)\citenamefont
  {Archidiacono}, \citenamefont {Gariazzo}, \citenamefont {Giunti},
  \citenamefont {Hannestad},\ and\ \citenamefont
  {Tram}}]{Archidiacono:2020yey}%
  \BibitemOpen
  \bibfield  {author} {\bibinfo {author} {\bibfnamefont {M.}~\bibnamefont
  {Archidiacono}}, \bibinfo {author} {\bibfnamefont {S.}~\bibnamefont
  {Gariazzo}}, \bibinfo {author} {\bibfnamefont {C.}~\bibnamefont {Giunti}},
  \bibinfo {author} {\bibfnamefont {S.}~\bibnamefont {Hannestad}}, \ and\
  \bibinfo {author} {\bibfnamefont {T.}~\bibnamefont {Tram}},\ }\href@noop {}
  {\  (\bibinfo {year} {2020})},\ \Eprint {http://arxiv.org/abs/2006.12885}
  {arXiv:2006.12885 [astro-ph.CO]} \BibitemShut {NoStop}%
\bibitem [{\citenamefont {Parker}\ and\ \citenamefont
  {Raval}(2000)}]{Parker:2000pr}%
  \BibitemOpen
  \bibfield  {author} {\bibinfo {author} {\bibfnamefont {L.}~\bibnamefont
  {Parker}}\ and\ \bibinfo {author} {\bibfnamefont {A.}~\bibnamefont {Raval}},\
  }\href {\doibase 10.1103/PhysRevD.62.083503} {\bibfield  {journal} {\bibinfo
  {journal} {Phys. Rev. D}\ }\textbf {\bibinfo {volume} {62}},\ \bibinfo
  {pages} {083503} (\bibinfo {year} {2000})},\ \bibinfo {note} {[Erratum:
  Phys.Rev.D 67, 029903 (2003)]},\ \Eprint {http://arxiv.org/abs/gr-qc/0003103}
  {arXiv:gr-qc/0003103} \BibitemShut {NoStop}%
\bibitem [{\citenamefont {Parker}\ and\ \citenamefont
  {Vanzella}(2004)}]{Parker:2003as}%
  \BibitemOpen
  \bibfield  {author} {\bibinfo {author} {\bibfnamefont {L.}~\bibnamefont
  {Parker}}\ and\ \bibinfo {author} {\bibfnamefont {D.~A.}\ \bibnamefont
  {Vanzella}},\ }\href {\doibase 10.1103/PhysRevD.69.104009} {\bibfield
  {journal} {\bibinfo  {journal} {Phys. Rev. D}\ }\textbf {\bibinfo {volume}
  {69}},\ \bibinfo {pages} {104009} (\bibinfo {year} {2004})},\ \Eprint
  {http://arxiv.org/abs/gr-qc/0312108} {arXiv:gr-qc/0312108} \BibitemShut
  {NoStop}%
\bibitem [{\citenamefont {Caldwell}\ \emph {et~al.}(2006)\citenamefont
  {Caldwell}, \citenamefont {Komp}, \citenamefont {Parker},\ and\ \citenamefont
  {Vanzella}}]{Caldwell:2005xb}%
  \BibitemOpen
  \bibfield  {author} {\bibinfo {author} {\bibfnamefont {R.~R.}\ \bibnamefont
  {Caldwell}}, \bibinfo {author} {\bibfnamefont {W.}~\bibnamefont {Komp}},
  \bibinfo {author} {\bibfnamefont {L.}~\bibnamefont {Parker}}, \ and\ \bibinfo
  {author} {\bibfnamefont {D.~A.}\ \bibnamefont {Vanzella}},\ }\href {\doibase
  10.1103/PhysRevD.73.023513} {\bibfield  {journal} {\bibinfo  {journal} {Phys.
  Rev. D}\ }\textbf {\bibinfo {volume} {73}},\ \bibinfo {pages} {023513}
  (\bibinfo {year} {2006})},\ \Eprint {http://arxiv.org/abs/astro-ph/0507622}
  {arXiv:astro-ph/0507622} \BibitemShut {NoStop}%
\bibitem [{\citenamefont {Starobinsky}(1987)}]{Starobinsky:1980te}%
  \BibitemOpen
  \bibfield  {author} {\bibinfo {author} {\bibfnamefont {A.~A.}\ \bibnamefont
  {Starobinsky}},\ }\href {\doibase 10.1016/0370-2693(80)90670-X} {\bibfield
  {journal} {\bibinfo  {journal} {Adv. Ser. Astrophys. Cosmol.}\ }\textbf
  {\bibinfo {volume} {3}},\ \bibinfo {pages} {130} (\bibinfo {year}
  {1987})}\BibitemShut {NoStop}%
\bibitem [{\citenamefont {Sakharov}(1991)}]{Sakharov:1967pk}%
  \BibitemOpen
  \bibfield  {author} {\bibinfo {author} {\bibfnamefont {A.}~\bibnamefont
  {Sakharov}},\ }\href {\doibase 10.1070/PU1991v034n05ABEH002498} {\bibfield
  {journal} {\bibinfo  {journal} {Usp. Fiz. Nauk}\ }\textbf {\bibinfo {volume}
  {161}},\ \bibinfo {pages} {64} (\bibinfo {year} {1991})}\BibitemShut
  {NoStop}%
\bibitem [{\citenamefont {Aghanim}\ \emph {et~al.}(2019)\citenamefont {Aghanim}
  \emph {et~al.}}]{Aghanim:2019ame}%
  \BibitemOpen
  \bibfield  {author} {\bibinfo {author} {\bibfnamefont {N.}~\bibnamefont
  {Aghanim}} \emph {et~al.} (\bibinfo {collaboration} {Planck}),\ }\href@noop
  {} {\  (\bibinfo {year} {2019})},\ \Eprint {http://arxiv.org/abs/1907.12875}
  {arXiv:1907.12875 [astro-ph.CO]} \BibitemShut {NoStop}%
\bibitem [{\citenamefont {Aghanim}\ \emph
  {et~al.}(2018{\natexlab{b}})\citenamefont {Aghanim} \emph
  {et~al.}}]{Aghanim:2018oex}%
  \BibitemOpen
  \bibfield  {author} {\bibinfo {author} {\bibfnamefont {N.}~\bibnamefont
  {Aghanim}} \emph {et~al.} (\bibinfo {collaboration} {Planck}),\ }\href@noop
  {} {\  (\bibinfo {year} {2018}{\natexlab{b}})},\ \Eprint
  {http://arxiv.org/abs/1807.06210} {arXiv:1807.06210 [astro-ph.CO]}
  \BibitemShut {NoStop}%
\bibitem [{\citenamefont {Beutler}\ \emph {et~al.}(2011)\citenamefont
  {Beutler}, \citenamefont {Blake}, \citenamefont {Colless}, \citenamefont
  {Jones}, \citenamefont {Staveley-Smith}, \citenamefont {Campbell},
  \citenamefont {Parker}, \citenamefont {Saunders},\ and\ \citenamefont
  {Watson}}]{Beutler:2011hx}%
  \BibitemOpen
  \bibfield  {author} {\bibinfo {author} {\bibfnamefont {F.}~\bibnamefont
  {Beutler}}, \bibinfo {author} {\bibfnamefont {C.}~\bibnamefont {Blake}},
  \bibinfo {author} {\bibfnamefont {M.}~\bibnamefont {Colless}}, \bibinfo
  {author} {\bibfnamefont {D.}~\bibnamefont {Jones}}, \bibinfo {author}
  {\bibfnamefont {L.}~\bibnamefont {Staveley-Smith}}, \bibinfo {author}
  {\bibfnamefont {L.}~\bibnamefont {Campbell}}, \bibinfo {author}
  {\bibfnamefont {Q.}~\bibnamefont {Parker}}, \bibinfo {author} {\bibfnamefont
  {W.}~\bibnamefont {Saunders}}, \ and\ \bibinfo {author} {\bibfnamefont
  {F.}~\bibnamefont {Watson}},\ }\href {\doibase
  10.1111/j.1365-2966.2011.19250.x} {\bibfield  {journal} {\bibinfo  {journal}
  {Mon. Not. Roy. Astron. Soc.}\ }\textbf {\bibinfo {volume} {416}},\ \bibinfo
  {pages} {3017} (\bibinfo {year} {2011})},\ \Eprint
  {http://arxiv.org/abs/1106.3366} {arXiv:1106.3366 [astro-ph.CO]} \BibitemShut
  {NoStop}%
\bibitem [{\citenamefont {Ross}\ \emph {et~al.}(2015)\citenamefont {Ross},
  \citenamefont {Samushia}, \citenamefont {Howlett}, \citenamefont {Percival},
  \citenamefont {Burden},\ and\ \citenamefont {Manera}}]{Ross:2014qpa}%
  \BibitemOpen
  \bibfield  {author} {\bibinfo {author} {\bibfnamefont {A.~J.}\ \bibnamefont
  {Ross}}, \bibinfo {author} {\bibfnamefont {L.}~\bibnamefont {Samushia}},
  \bibinfo {author} {\bibfnamefont {C.}~\bibnamefont {Howlett}}, \bibinfo
  {author} {\bibfnamefont {W.~J.}\ \bibnamefont {Percival}}, \bibinfo {author}
  {\bibfnamefont {A.}~\bibnamefont {Burden}}, \ and\ \bibinfo {author}
  {\bibfnamefont {M.}~\bibnamefont {Manera}},\ }\href {\doibase
  10.1093/mnras/stv154} {\bibfield  {journal} {\bibinfo  {journal} {Mon. Not.
  Roy. Astron. Soc.}\ }\textbf {\bibinfo {volume} {449}},\ \bibinfo {pages}
  {835} (\bibinfo {year} {2015})},\ \Eprint {http://arxiv.org/abs/1409.3242}
  {arXiv:1409.3242 [astro-ph.CO]} \BibitemShut {NoStop}%
\bibitem [{\citenamefont {Alam}\ \emph {et~al.}(2017)\citenamefont {Alam} \emph
  {et~al.}}]{Alam:2016hwk}%
  \BibitemOpen
  \bibfield  {author} {\bibinfo {author} {\bibfnamefont {S.}~\bibnamefont
  {Alam}} \emph {et~al.} (\bibinfo {collaboration} {BOSS}),\ }\href {\doibase
  10.1093/mnras/stx721} {\bibfield  {journal} {\bibinfo  {journal} {Mon. Not.
  Roy. Astron. Soc.}\ }\textbf {\bibinfo {volume} {470}},\ \bibinfo {pages}
  {2617} (\bibinfo {year} {2017})},\ \Eprint {http://arxiv.org/abs/1607.03155}
  {arXiv:1607.03155 [astro-ph.CO]} \BibitemShut {NoStop}%
\bibitem [{\citenamefont {Scolnic}\ \emph {et~al.}(2018)\citenamefont {Scolnic}
  \emph {et~al.}}]{Scolnic:2017caz}%
  \BibitemOpen
  \bibfield  {author} {\bibinfo {author} {\bibfnamefont {D.}~\bibnamefont
  {Scolnic}} \emph {et~al.},\ }\href {\doibase 10.3847/1538-4357/aab9bb}
  {\bibfield  {journal} {\bibinfo  {journal} {Astrophys. J.}\ }\textbf
  {\bibinfo {volume} {859}},\ \bibinfo {pages} {101} (\bibinfo {year}
  {2018})},\ \Eprint {http://arxiv.org/abs/1710.00845} {arXiv:1710.00845
  [astro-ph.CO]} \BibitemShut {NoStop}%
\bibitem [{\citenamefont {Lewis}\ and\ \citenamefont
  {Bridle}(2002)}]{Lewis:2002ah}%
  \BibitemOpen
  \bibfield  {author} {\bibinfo {author} {\bibfnamefont {A.}~\bibnamefont
  {Lewis}}\ and\ \bibinfo {author} {\bibfnamefont {S.}~\bibnamefont {Bridle}},\
  }\href {\doibase 10.1103/PhysRevD.66.103511} {\bibfield  {journal} {\bibinfo
  {journal} {Phys. Rev. D}\ }\textbf {\bibinfo {volume} {66}},\ \bibinfo
  {pages} {103511} (\bibinfo {year} {2002})},\ \Eprint
  {http://arxiv.org/abs/astro-ph/0205436} {arXiv:astro-ph/0205436} \BibitemShut
  {NoStop}%
\bibitem [{\citenamefont {Gelman}\ and\ \citenamefont
  {Rubin}(1992)}]{Gelman:1992zz}%
  \BibitemOpen
  \bibfield  {author} {\bibinfo {author} {\bibfnamefont {A.}~\bibnamefont
  {Gelman}}\ and\ \bibinfo {author} {\bibfnamefont {D.~B.}\ \bibnamefont
  {Rubin}},\ }\href {\doibase 10.1214/ss/1177011136} {\bibfield  {journal}
  {\bibinfo  {journal} {Statist. Sci.}\ }\textbf {\bibinfo {volume} {7}},\
  \bibinfo {pages} {457} (\bibinfo {year} {1992})}\BibitemShut {NoStop}%
\bibitem [{\citenamefont {Lewis}(2013)}]{Lewis:2013hha}%
  \BibitemOpen
  \bibfield  {author} {\bibinfo {author} {\bibfnamefont {A.}~\bibnamefont
  {Lewis}},\ }\href {\doibase 10.1103/PhysRevD.87.103529} {\bibfield  {journal}
  {\bibinfo  {journal} {Phys. Rev. D}\ }\textbf {\bibinfo {volume} {87}},\
  \bibinfo {pages} {103529} (\bibinfo {year} {2013})},\ \Eprint
  {http://arxiv.org/abs/1304.4473} {arXiv:1304.4473 [astro-ph.CO]} \BibitemShut
  {NoStop}%
\bibitem [{\citenamefont {Rubin}\ \emph {et~al.}(2009)\citenamefont {Rubin}
  \emph {et~al.}}]{Rubin:2008wq}%
  \BibitemOpen
  \bibfield  {author} {\bibinfo {author} {\bibfnamefont {D.}~\bibnamefont
  {Rubin}} \emph {et~al.},\ }\href {\doibase 10.1088/0004-637X/695/1/391}
  {\bibfield  {journal} {\bibinfo  {journal} {Astrophys. J.}\ }\textbf
  {\bibinfo {volume} {695}},\ \bibinfo {pages} {391} (\bibinfo {year}
  {2009})},\ \Eprint {http://arxiv.org/abs/0807.1108} {arXiv:0807.1108
  [astro-ph]} \BibitemShut {NoStop}%
\bibitem [{\citenamefont {Handley}(2019)}]{Handley:2019tkm}%
  \BibitemOpen
  \bibfield  {author} {\bibinfo {author} {\bibfnamefont {W.}~\bibnamefont
  {Handley}},\ }\href@noop {} {\  (\bibinfo {year} {2019})},\ \Eprint
  {http://arxiv.org/abs/1908.09139} {arXiv:1908.09139 [astro-ph.CO]}
  \BibitemShut {NoStop}%
\bibitem [{\citenamefont {Di~Valentino}\ \emph
  {et~al.}(2019{\natexlab{c}})\citenamefont {Di~Valentino}, \citenamefont
  {Melchiorri},\ and\ \citenamefont {Silk}}]{DiValentino:2019qzk}%
  \BibitemOpen
  \bibfield  {author} {\bibinfo {author} {\bibfnamefont {E.}~\bibnamefont
  {Di~Valentino}}, \bibinfo {author} {\bibfnamefont {A.}~\bibnamefont
  {Melchiorri}}, \ and\ \bibinfo {author} {\bibfnamefont {J.}~\bibnamefont
  {Silk}},\ }\href {\doibase 10.1038/s41550-019-0906-9} {\bibfield  {journal}
  {\bibinfo  {journal} {Nature Astron.}\ }\textbf {\bibinfo {volume} {4}},\
  \bibinfo {pages} {196} (\bibinfo {year} {2019}{\natexlab{c}})},\ \Eprint
  {http://arxiv.org/abs/1911.02087} {arXiv:1911.02087 [astro-ph.CO]}
  \BibitemShut {NoStop}%
\bibitem [{\citenamefont {Efstathiou}\ and\ \citenamefont
  {Gratton}(2020)}]{Efstathiou:2020wem}%
  \BibitemOpen
  \bibfield  {author} {\bibinfo {author} {\bibfnamefont {G.}~\bibnamefont
  {Efstathiou}}\ and\ \bibinfo {author} {\bibfnamefont {S.}~\bibnamefont
  {Gratton}},\ }\href {\doibase 10.1093/mnrasl/slaa093} {\bibfield  {journal}
  {\bibinfo  {journal} {Mon. Not. Roy. Astron. Soc.}\ }\textbf {\bibinfo
  {volume} {496}},\ \bibinfo {pages} {L91} (\bibinfo {year} {2020})},\ \Eprint
  {http://arxiv.org/abs/2002.06892} {arXiv:2002.06892 [astro-ph.CO]}
  \BibitemShut {NoStop}%
\bibitem [{\citenamefont {Liao}\ \emph {et~al.}(2020)\citenamefont {Liao},
  \citenamefont {Shafieloo}, \citenamefont {Keeley},\ and\ \citenamefont
  {Linder}}]{Liao:2020zko}%
  \BibitemOpen
  \bibfield  {author} {\bibinfo {author} {\bibfnamefont {K.}~\bibnamefont
  {Liao}}, \bibinfo {author} {\bibfnamefont {A.}~\bibnamefont {Shafieloo}},
  \bibinfo {author} {\bibfnamefont {R.~E.}\ \bibnamefont {Keeley}}, \ and\
  \bibinfo {author} {\bibfnamefont {E.~V.}\ \bibnamefont {Linder}},\ }\href
  {\doibase 10.3847/2041-8213/ab8dbb} {\bibfield  {journal} {\bibinfo
  {journal} {Astrophys. J.}\ }\textbf {\bibinfo {volume} {895}},\ \bibinfo
  {pages} {L29} (\bibinfo {year} {2020})},\ \Eprint
  {http://arxiv.org/abs/2002.10605} {arXiv:2002.10605 [astro-ph.CO]}
  \BibitemShut {NoStop}%
\bibitem [{\citenamefont {Di~Valentino}\ \emph
  {et~al.}(2020{\natexlab{b}})\citenamefont {Di~Valentino}, \citenamefont
  {Melchiorri},\ and\ \citenamefont {Silk}}]{DiValentino:2020hov}%
  \BibitemOpen
  \bibfield  {author} {\bibinfo {author} {\bibfnamefont {E.}~\bibnamefont
  {Di~Valentino}}, \bibinfo {author} {\bibfnamefont {A.}~\bibnamefont
  {Melchiorri}}, \ and\ \bibinfo {author} {\bibfnamefont {J.}~\bibnamefont
  {Silk}},\ }\href@noop {} {\  (\bibinfo {year} {2020}{\natexlab{b}})},\
  \Eprint {http://arxiv.org/abs/2003.04935} {arXiv:2003.04935 [astro-ph.CO]}
  \BibitemShut {NoStop}%
\bibitem [{\citenamefont {Linder}(2017)}]{Linder:2016xer}%
  \BibitemOpen
  \bibfield  {author} {\bibinfo {author} {\bibfnamefont {E.~V.}\ \bibnamefont
  {Linder}},\ }\href {\doibase 10.1016/j.astropartphys.2016.11.002} {\bibfield
  {journal} {\bibinfo  {journal} {Astropart. Phys.}\ }\textbf {\bibinfo
  {volume} {86}},\ \bibinfo {pages} {41} (\bibinfo {year} {2017})},\ \Eprint
  {http://arxiv.org/abs/1610.05321} {arXiv:1610.05321 [astro-ph.CO]}
  \BibitemShut {NoStop}%
\bibitem [{\citenamefont {{Linder}}(2005)}]{2005PhRvD..72d3529L}%
  \BibitemOpen
  \bibfield  {author} {\bibinfo {author} {\bibfnamefont {E.~V.}\ \bibnamefont
  {{Linder}}},\ }\href {\doibase 10.1103/PhysRevD.72.043529} {\bibfield
  {journal} {\bibinfo  {journal} {\prd}\ }\textbf {\bibinfo {volume} {72}},\
  \bibinfo {eid} {043529} (\bibinfo {year} {2005})},\ \Eprint
  {http://arxiv.org/abs/astro-ph/0507263} {arXiv:astro-ph/0507263 [astro-ph]}
  \BibitemShut {NoStop}%
\bibitem [{\citenamefont {{Linder}}\ and\ \citenamefont
  {{Cahn}}(2007)}]{2007APh....28..481L}%
  \BibitemOpen
  \bibfield  {author} {\bibinfo {author} {\bibfnamefont {E.~V.}\ \bibnamefont
  {{Linder}}}\ and\ \bibinfo {author} {\bibfnamefont {R.~N.}\ \bibnamefont
  {{Cahn}}},\ }\href {\doibase 10.1016/j.astropartphys.2007.09.003} {\bibfield
  {journal} {\bibinfo  {journal} {Astroparticle Physics}\ }\textbf {\bibinfo
  {volume} {28}},\ \bibinfo {pages} {481} (\bibinfo {year} {2007})},\ \Eprint
  {http://arxiv.org/abs/astro-ph/0701317} {arXiv:astro-ph/0701317 [astro-ph]}
  \BibitemShut {NoStop}%
\bibitem [{\citenamefont {{Wang}}\ and\ \citenamefont
  {{Steinhardt}}(1998)}]{wangs}%
  \BibitemOpen
  \bibfield  {author} {\bibinfo {author} {\bibfnamefont {L.}~\bibnamefont
  {{Wang}}}\ and\ \bibinfo {author} {\bibfnamefont {P.~J.}\ \bibnamefont
  {{Steinhardt}}},\ }\href {\doibase 10.1086/306436} {\bibfield  {journal}
  {\bibinfo  {journal} {\apj}\ }\textbf {\bibinfo {volume} {508}},\ \bibinfo
  {pages} {483} (\bibinfo {year} {1998})},\ \Eprint
  {http://arxiv.org/abs/astro-ph/9804015} {arXiv:astro-ph/9804015 [astro-ph]}
  \BibitemShut {NoStop}%
\bibitem [{\citenamefont {Di~Valentino}\ \emph {et~al.}(2017)\citenamefont
  {Di~Valentino}, \citenamefont {Melchiorri}, \citenamefont {Linder},\ and\
  \citenamefont {Silk}}]{DiValentino:2017zyq}%
  \BibitemOpen
  \bibfield  {author} {\bibinfo {author} {\bibfnamefont {E.}~\bibnamefont
  {Di~Valentino}}, \bibinfo {author} {\bibfnamefont {A.}~\bibnamefont
  {Melchiorri}}, \bibinfo {author} {\bibfnamefont {E.~V.}\ \bibnamefont
  {Linder}}, \ and\ \bibinfo {author} {\bibfnamefont {J.}~\bibnamefont
  {Silk}},\ }\href {\doibase 10.1103/PhysRevD.96.023523} {\bibfield  {journal}
  {\bibinfo  {journal} {Phys. Rev.}\ }\textbf {\bibinfo {volume} {D96}},\
  \bibinfo {pages} {023523} (\bibinfo {year} {2017})},\ \Eprint
  {http://arxiv.org/abs/1704.00762} {arXiv:1704.00762 [astro-ph.CO]}
  \BibitemShut {NoStop}%
\end{thebibliography}%
\end{document}